\documentclass{sig-alternate-05-2015}
\usepackage{graphicx}  %Required
\usepackage{fancyhdr}
\usepackage{lastpage}
\usepackage{tocloft}
\usepackage{lmodern}
\usepackage[utf8]{inputenc}
\usepackage{csvsimple}
\usepackage{amsmath}
\usepackage{mathptmx}
\usepackage{eqnarray}
\usepackage{stmaryrd}
\usepackage{balance}
\usepackage{color}
\usepackage{booktabs}
\usepackage{etoolbox}
\usepackage{subfig}
\usepackage{float}
\usepackage{dsfont}
\usepackage{microtype}
\usepackage{xcolor,colortbl}
\makeatletter
\def\@copyrightspace{\relax}
\makeatother

\begin{document}

% Copyright
\setcopyright{acmcopyright}
%\setcopyright{acmlicensed}
%\setcopyright{rightsretained}
%\setcopyright{usgov}
%\setcopyright{usgovmixed}
%\setcopyright{cagov}
%\setcopyright{cagovmixed}

% DOI
\doi{10.475/123_4}

% ISBN
\isbn{123-4567-24-567/08/06}

%Conference
\conferenceinfo{PLDI '13}{June 16--19, 2013, Seattle, WA, USA}

\acmPrice{\$15.00}

%
% --- Author Metadata here ---
\conferenceinfo{WOODSTOCK}{'97 El Paso, Texas USA}
%\CopyrightYear{2007} % Allows default copyright year (20XX) to be over-ridden - IF NEED BE.
%\crdata{0-12345-67-8/90/01}  % Allows default copyright data (0-89791-88-6/97/05) to be over-ridden - IF NEED BE.
% --- End of Author Metadata ---

\title{20 Years of Mobility Modeling \& Prediction:\\ Trends, Shortcomings \& Perspectives}

%\subtitle{[Extended Abstract]
%\titlenote{A full version of this paper is available as
%\textit{Author's Guide to Preparing ACM SIG Proceedings Using
%\LaTeX$2_\epsilon$\ and BibTeX} at

\numberofauthors{1} %  in this sample file, there are a *total*

\author{
% 1st. author
\alignauthor
Vaibhav Kulkarni and Beno\^{i}t Garbinato\\
\vspace{6px}
\email{\{firstname.lastname\}@unil.ch}\\
\affaddr{Distributed Object Programming Laboratory}\\
\affaddr{University of Lausanne, Switzerland}\\
}

\makeatletter

\maketitle

\begin{abstract}

\begin{figure*}[htp!]
\centering
\includegraphics[width=\textwidth]{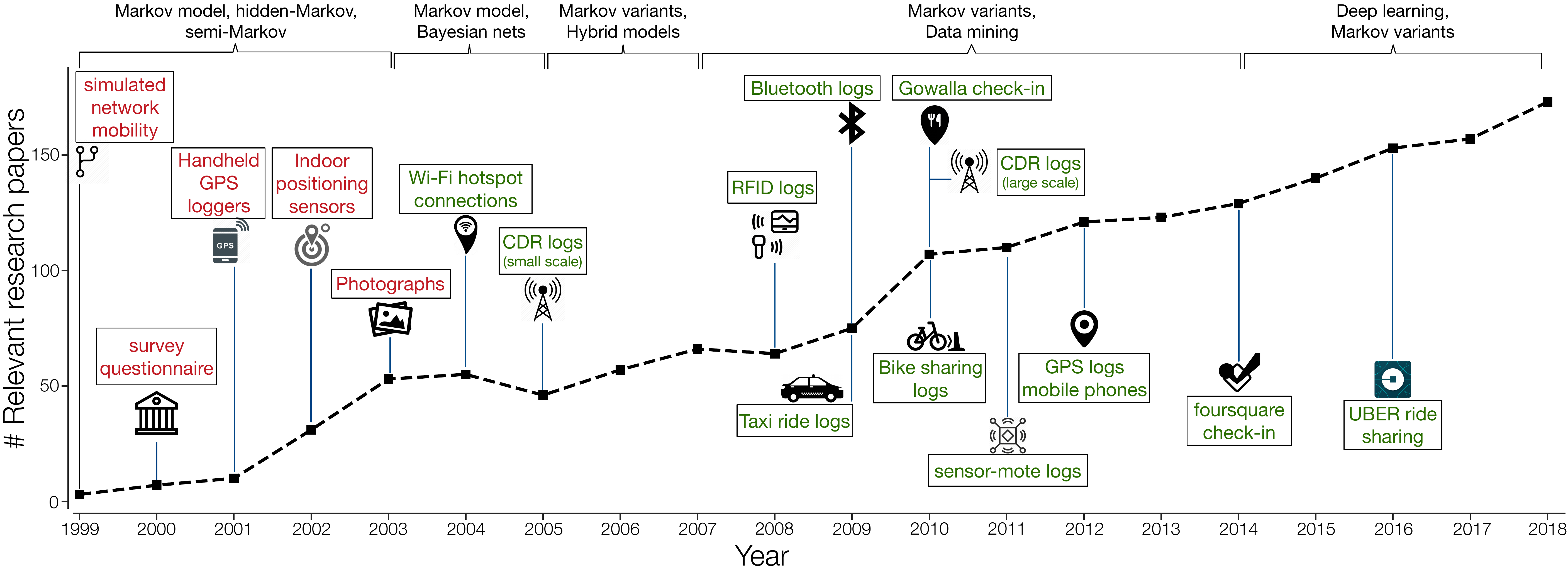}
 \caption{Mobility modeling and prediction: 20 years in review. The figure presents a summary of the review, highlighting the key techniques dominant (in terms of number of papers) in the respective era and the dataset (private datasets in red, public in green) driving this research.}
 \label{fig:mobility_mod_trend}
\end{figure*}

In this paper, we present a comprehensive survey of human-mobility modeling based on 1680 articles published between 1999 and 2019, which can serve as a roadmap for research and practice in this area.
Mobility modeling research has accelerated the advancement of several fields of studies such as urban planning, epidemic modeling, traffic engineering and contributed to the development of location-based services. 
However, while the application of mobility models in different domains has increased, the credibility of the research results has decreased.  
We highlight two significant shortfalls commonly observed in our reviewed studies: (1) data-agnostic model selection resulting in a poor tradeoff between accuracy vs. complexity, and (2) failure to identify the source of empirical gains, due to adoption of inaccurate validation methodologies. 
We also observe troubling trends with respect to application of Markov model variants for modeling mobility, despite the questionable association of Markov processes and human-mobility dynamics.
To this end, we propose a data-driven mobility-modeling framework that quantifies the characteristics of a dataset based on four mobility meta-attributes, in order to select the most appropriate prediction algorithm.  
Experimental evaluations on three real-world mobility datasets based on a rigorous validation methodology demonstrate our frameworks ability to correctly analyze the model accuracy vs. complexity tradeoff.   
We offer these results to the community along with the tools and the literature meta-data in order to improve the reliability and credibility of human mobility modeling research. 

\end{abstract}




% We no longer use \terms command


\section{Introduction}
\label{sec:introduction}

Over the last two decades, we have seen a large number of studies on human mobility modeling and prediction by the Geographic Information Systems (GIS) community.
This testifies of the importance of mobility prediction in context-aware systems where a user's future location is used to seamlessly trigger service execution. 
These systems span services such as ride sharing, traffic prediction, point of interest recommendation, resource/urban planning, and network optimization among others.
Given the sensitive nature of mobility trajectories and the enforcement of binding privacy regulations, privacy-preserving modeling approaches such as federated learning~\cite{smith2017federated} and Google Rappor are advancing.
In such approaches, the process of model training and updating is executed locally on resource-constrained smartphones~\cite{smith2017federated}.
A heuristically driven framework that analyzes the model performance vs. complexity trade-off for algorithmic selection is therefore essential.

To understand the current methodologies driving prediction-model selection and performance validation strategies, we performed a systematic literature review spanning last two decades amounting to 1680 articles.
Based on the reviewed literature, mobility modeling an be defined as the process of estimating the probability distribution over an individual's future movement by minimizing the negative log-likelihood over the currently known user trajectory.
Research in this domain can thus be classified in three distinct categories: (1) theoretical modeling of mobility dynamics, (2) quantifying the uncertainty in next-place prediction, and (3) leveraging stochastic optimization algorithms to model human-mobility and benchmark the next-place forecasting capability.  
This paper focuses on the third category, where the stochastic approaches applied for constructing the next-place forecasting model fall into three categories: (1) Markov model variants (2) data mining techniques, and (3) neural network architectures.
In order to validate the model's predictive performance, several types of datasets are used in these works that either contain GPS trajectories of pedestrians, recurrent WiFi connections, Bluetooth records or social network check-ins.

Despite the large number of studies, it is not trivial either to objectively compare cross-model performance, nor to identify the source of the empirical gains provided by the proposed models.
This difficulty stems from the fact that each modeling approach is implemented with distinct search heuristics which introduces a range of inductive biases~\cite{madala1994inductive}.
This results in delivering different performance depending upon the dataset attributes~\cite{Peng2002ImprovedDC}. 
We find that for a majority of instances, empirical gains predominantly stem from erroneous validation methodology and selection of datasets with opportune characteristics rather than modeling/architectural amendments.
Furthermore, the availability of several datasets and no explicit procedure to perform mobility-prediction validation makes it challenging to determine the appropriate learning algorithm.
We observe that 68\% of the reviewed work relies on variants of Markov models despite the unclear association between Markov processes and human mobility dynamics~\cite{kulkarni2019examining}.
As presented in Figure~\ref{fig:mobility_mod_trend}, this trend declines to a certain extent after the onset of deep learning and the availability of the associated implementation frameworks.

Knowing that there is no single algorithm that can perform uniformly over all the modeling tasks on different datasets (no free lunch theorem)~\cite{wolpert1996lack}, it is essential to select the appropriate mobility-modeling algorithm depending upon the dataset characteristics.
Meta-learning is a bias minimization perspective on learning algorithm selection, by accumulating meta-knowledge about the data~\cite{pfahringer2000meta}.
Such an approach considers statistics inspired measures, such as skewness, entropy and autocorrelation, as a source for the definition of dataset characteristics.
These characteristics are known as meta-attributes. 
Correctly estimating them speeds up and improves the mobility modeling pipeline design by achieving faster convergence, optimal local minimum and discernible models. 
We find a lack of research directed towards estimation of these meta-attributes for human mobility that can act as metrics characterizing the mobility datasets.
To this end, we experimentally show the importance of selecting prediction models based on the meta-attributes and highlight the consequences of adopting misleading validation methodologies. 
Relying on robust cross-validation approaches for sequential data is also essential to present a fair contribution of a model and to perform intuitive comparison with other models.
To this end, our paper makes contributions on the following fronts:

\begin{itemize}

	\item Through a large-scale literature survey, we systematize knowledge on mobility-modeling research and provide our insights on how this research should be conducted and which challenges it should address. We offer the tools and the literature meta-data to the community with the hope of improving the credibility of mobility modeling research.~\footnote{Link: {\url{https://bit.ly/2HRZGk5}}}
	
	\item We propose four meta-attributes to quantify mobility dataset characteristics grounded on statistical and information theoretic primitives: (1) average length of an individual trajectory, (2) number of points of interest, (3) number of points of interests interacting with each other, and (4) distance between the interactions of points of interest.
	
	\item Based on the meta-attributes, we present a mobility modeling framework to assist in selecting an appropriate learning algorithm. We then present a validation methodology to perform performance assessment of mobility prediction models allowing for a fair comparison and facilitating the process of incremental improvements. We evaluate the performance of our approach on three mobility datasets and discuss the associated accuracy vs. complexity trade-offs.

\end{itemize}

The remainder of the paper is organized as follows. 
Section~\ref{sec:survey_methodology} presents the survey methodology, followed by the survey findings in Section~\ref{sec:review_summary}.
Section~\ref{sec:shortcomings} highlights the shortcomings of the current mobility modeling approaches. 
In Section~\ref{sec:data_driven_approach}, we present our data-driven mobility modeling framework, followed by the results and discussion in Section~\ref{sec:results_discussion}.
We conclude the paper in Section~\ref{sec:conclusion}.

\begin{figure*}[t!]
    \centering
    \includegraphics[width=0.83\textwidth]{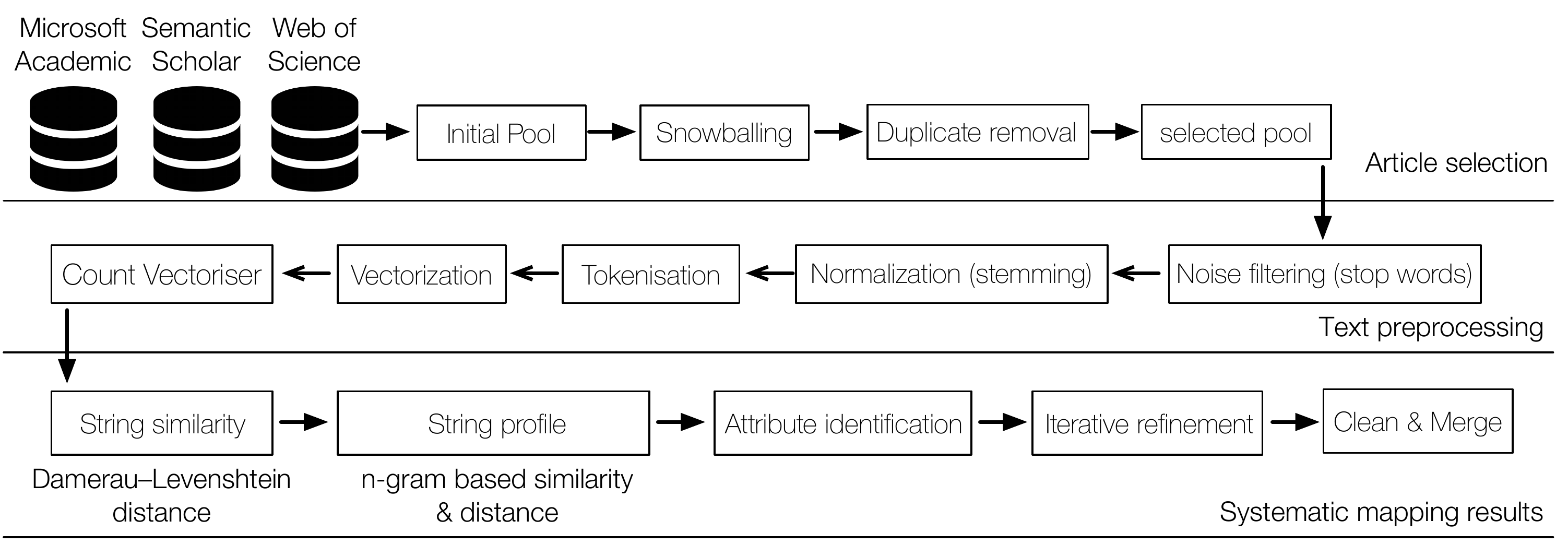}
    \caption{Overview of the search \& data extraction strategy, study selection and quality assessment methodology.}
    \label{fig:scrapping_refine}
    \vspace{-8px}
\end{figure*}

\section{Survey Methodology}
\label{sec:survey_methodology}

The increasing number of articles published on human mobility-modeling and prediction testifies to the growing interest from researchers and practitioners.
Figure~\ref{fig:mobility_mod_trend} shows the most significant prediction models and the datasets used for mobility prediction for a given span of year(s).
The goal of the systematic literature review is to classify the state-of-the-art in the area of human mobility modeling and answer the following research questions: (1) which techniques are used for constructing the next-place prediction models?, (2) which methodologies are adopted to validate the performance of the models?, and (3) which datasets are used to perform the performance quantification?. 

Despite the large number of hits on these topics, there is a dearth of coherent understanding on what kind of studies have been conducted under the term {\it{human mobility modeling}}, with which methods, what kind of results they yield, and under which circumstances. 
Thus in this section, we describe the methodology we adopted for source selection, search keywords, followed by quality assessment and application of article inclusion, exclusion criteria. 
Schreckenberger et al.~\cite{schreckenberger2018next} perform a similar survey spanning from 2013 to 2018 and present the descriptive statistics with regards to the data types and techniques used. 
On the contrary, our study spans the last 20 years and we not only provide the overview, but we perform an in-depth study on the impact of datasets, techniques, validation methodologies and experimentally validate the problematic trends,
In addition, we also conduct experiments on real-world mobility datasets to expose the pitfalls in the currently adopted techniques and propose amendments.   

\subsection{Search Strategy \& Article Selection}

To find relevant studies, we searched three major academic article search engines: Web of Science, Microsoft Academic Search and Semantic Scholar.  
Three distinct platforms were chosen to ensure result completeness, as a single platform does not cover all major publisher venues.
The search terms included, {\it{human-mobility prediction, human-mobility modeling, next place forecasting, Predicting Significant locations}} among others.
The search terms were used for the title, abstract and keywords.
These search domains consisted of {\it{mobile and ubiquitous computing, geographic information systems and knowledge discovery and data mining}}.
Furthermore, only the studies that were peer reviewed, published in an international venue, written in English language and that were electronically available were included.
Additional papers were identified by using the citation and reference list of a given paper to minimize the risk of omitting relevant studies.  
This step is according to the guidelines recommended by Keele et al.~\cite{keele2007guidelines} to conduct a systematic literature review and is known as selected snowballing. 
We selected the top five cited articles in the domain of human mobility prediction~\cite{Song2010LimitsOP, gambs2012next, Ashbrook2003UsingGT, Mathew2012PredictingFL, Monreale2009WhereNextAL} to bootstrap the snowballing procedure.
At this stage, studies in the pool were ready for application of inclusion/exclusion criteria. 
We complete list of keywords, search strings, domains and the associated scripts are published along with the literature meta-data.

\subsection{Preprocessing \& Quality Assessment}

To eliminate duplicates, the article titles were preprocessed by removing the stop words and performing normalization (stemming).
String similarity was estimated by using Damerau-Levenshtein distance, which measures the edit distance between two sequences. 
To calibrate the distance threshold, we manually assessed the top 20 top cited articles and labeled them according to their relevance.
The prediction technique used in each paper was identified by first normalizing the abstract followed by stop-words removal, tokenization, vectorization and  using a count vectorizer to analyze relative frequency of the tri-grams~\cite{bharti2017automatic}.
This is a simple statistical approach that analyzes the position of word in the abstract, the term frequency and inverse document frequency. 
The preprocessing and quality assessment pipeline is summarized in Figure~\ref{fig:scrapping_refine}.
This cleaned list of articles were used to perform the analysis presented in the next section of the paper.  

\begin{table*}[t!]
\centering
\resizebox{\textwidth}{!}{%
\begin{tabular}{lll}
\hline
\textbf{Approach} & \textbf{Dataset (\#participants, duration, type, location)} & \textbf{Validation methodology} \\ \hline
\rowcolor[HTML]{EFEFEF} 
Markov model~\cite{Ashbrook2001LearningSL} & 1 participant, 4 months, GPS traces & Not specified \\
semi-Markov model~\cite{chon2012evaluating} & 10 participants, unknown duration , GPS, GSM, WiFi & Not specified \\
\rowcolor[HTML]{EFEFEF} 
hidden Markov model~\cite{Mathew2012PredictingFL} & GeoLife (182 participants, 5 years, GPS traces) & Leave one out validation \\
mixed Makov model~\cite{Asahara2011PedestrianmovementPB} & Simulated data: 691 participants, 1hr;31 minutes, GPS & k-fold cross validation (k=10) \\
\rowcolor[HTML]{EFEFEF} 
mobility Markov chain~\cite{gambs2012next} &  GeoLife, synthetic dataset, private dataset (6 researchers) & Holdout validation, 50\% split train-test \\
extended mobility Markov chain~\cite{amirrudin2013user} & 1 participant, 54 weeks, CDR & Not specified \\
\rowcolor[HTML]{EFEFEF} 
variable-order Markov model~\cite{yan2013diversity} & GeoLife (182 participants, 5 years, GPS traces) & Holdout, Random selection, 90\%-10\% train-test \\
hidden semi-Markov model~\cite{Yu2003AHS} & Simulated data & No empirical validation \\
\rowcolor[HTML]{EFEFEF} 
hierarchical semi-Markov model~\cite{Baratchi2014AHH} &  GeoLife (182 participants, 5 years, GPS traces) & Holdout validation (no split \% specified) \\
spherical hidden-Markov model~\cite{zhu2018spherical} & Simulated data, Twitter dataset (geotagged tweets) & Random selection 70\%-30\% train-test split \\
\rowcolor[HTML]{EFEFEF}
Adaboost-Markov model~\cite{wang2019next} &  GeoLife (182 participants, 5 years, GPS traces) & Random selection 90\%-10\% train-test split \\ \hline
\end{tabular}%
}
\caption{Different variants of Markov models used to model human-mobility, datasets used to corroborate the model performance and the different types of cross-validation strategies applied.}
\label{tbl:markov_model_variants}
\vspace{-10px}
\end{table*}

\section{Survey Findings}
\label{sec:review_summary}

The first application of human mobility prediction was in the context of ad-hoc wireless networks by Gerla~\cite{Gerla1999IPv6FH}.
The knowledge about the user's next-location was used to anticipate topological changes and minimize the connectivity disruption caused by mobility.
Application of human mobility prediction in wireless networks became prominent after Su et al. in~\cite{Su2001MobilityPA}, proposed a location-aware routing scheme and demonstrated its effectiveness using simulations. 
The seminal work however, in the context of prediction location on road networks using GPS information, was made by Ashbrook et al.~\cite{Ashbrook2001LearningSL}. 
They first identified significant places (points of interest) by clustering the raw GPS trajectories and then built a Markov-based predictor to forecast the next significant place.  
In their next article, Ashbrook et al.~\cite{Ashbrook2003UsingGT} extended this approach to perform mobility prediction across a dataset having multiple users. 
Several approaches along these lines were presented in the subsequent years using geolocation datasets consisting of data-points from different mobile phone sensors~\cite{Patterson2003InferringHB, Soh2003QoSPI, Pathirana2003MobilityMA}.

The reviewed studies can be categorized based on the techniques applied for mobility prediction as follows: (1) Markov model variants, (2) neural network techniques, and (3) data-mining based approaches.
Table~\ref{tbl:markov_model_variants} presents a meta-summary of the Markov model variants, the dataset used to quantify the model performance and the validation methodology used.
A similar analysis based on neural network and data mining techniques is presented in Table~\ref{tbl:rnn_variants} and Table~\ref{tbl:data_mining} respectively. 
These studies have used several datasets differing with respect to the data type, number of users, collection duration and geographic regions.
We also highlight that in several cases the datasets are obtained privately from the telecommunication operators or generated synthetically using unspecified mobility simulators.

Accessibility of larger datasets prompted development and application of several variants of predictors based on Markov model such as hidden Markov model~\cite{Mathew2012PredictingFL}, mixed Markov model~\cite{Asahara2011PedestrianmovementPB}, semi-Markov model~\cite{chon2012evaluating}, hidden semi-Markov model~\cite{Yu2003AHS}, mobility Markov chain~\cite{gambs2012next}, extended mobility Markov chain~\cite{amirrudin2013user}, variable order Markov model~\cite{yang2014predicting}, hierarchical hidden Markov model\cite{Baratchi2014AHH} and spherical hidden Markov model~\cite{Baratchi2014AHH}.
Each of the variant claims to address and account for different aspects of mobility trajectories, such as missing data from some time intervals~\cite{Yu2003AHS}, correlation with the stay duration~\cite{chon2012evaluating}, memory requirement~\cite{yang2014predicting}, user behavioral characteristics~\cite{Asahara2011PedestrianmovementPB}, spatiotemporal associations with the path connecting the stay-points~\cite{Baratchi2014AHH}, semantic trace data~\cite{zhu2018spherical} or location specific characteristics~\cite{Mathew2012PredictingFL}.
Other approaches, such as non-parametric Bayesian model were also leveraged for construction of mobility-prediction models~\cite{jeong2015mobility}.
Here, particle filters and expectation-maximization techniques were applied to forecast the next place of the users.

After the prominence of big data in 2008, several data mining techniques, which explore periodic patterns and association rules, were employed for mobility prediction. 
Monreale et al.~\cite{Monreale2009WhereNextAL} define a trajectory as an ordered sequence of time-stamped locations and propose a modified version of the a-priori algorithm based on sequence analysis.
Such techniques do not generalize as well as state-space models, due to ignoring the notion of spatiotemporal distance.
Another type of approach falls in the category of template matching, where extracted features from time-stamped sequences are compared with pre-stored templates, using similarity search metrics such as dynamic time warping~\cite{Riedel2008RecognisingOS}.
Other variations such edit distance related metrics were also applied for template matching schemes~\cite{ossama2009similarity}.

After 2014, we observe an onset of application of neural network techniques for mobility prediction. 
The principle argument for applying deep learning networks is that the probabilistic models lack fine granularity in prediction and suffer from data sparsity problem.  
A recurrent neural network (RNN) based attention model for predicting the next place was proposed by Feng et al.~\cite{Feng2018DeepMovePH}.
This model is able to captures the multi-level periodicities present in human mobility.  
A bidirectional Long Short-Term Memory (LSTM) network architecture was proposed by Zhao et al.~\cite{Zhao2018OnPO} to predict the trip destination location. 
This network provides higher attention to locations having strong correlations with the destination due to the attention mechanism. 
A sequence to sequence (Seq2Seq) approach was applied by Karatzoglou et al.~\cite{Karatzoglou2018ASL} to human semantic trajectories, in order to improve prediction of semantically annotated trajectories.  
Al-Molegi et al.~\cite{AlMolegi2016STFRNNST} proposed a space time feature-based RNN for predicting next user movement.
The model operates by conditioning the RNN inputs based on the spatial features and temporal elements present in the trajectory. 
Liu et al.~\cite{liu2016predicting} proposed an extension of the vanilla-RNN model, where each layer is upon different time intervals and distance specific transition matrices for distinct geographical distances.

\begin{table}[t!]
\centering
\resizebox{0.47\textwidth}{!}{%
\begin{tabular}{lll}
\hline
\textbf{Approach} & \textbf{Dataset} & \textbf{Validation methodology} \\ \hline
\rowcolor[HTML]{EFEFEF} 
Attentional RNN~\cite{Feng2018DeepMovePH} & {\color[HTML]{333333} \begin{tabular}[c]{@{}l@{}}Foursquare Check-ins\\ Mobile apps (private)\\ CDR data (private)\end{tabular}} & \begin{tabular}[c]{@{}l@{}}Holdout (80\%-20\% split)\end{tabular} \\
Bi-dir. LSTM~\cite{Zhao2018OnPO} & \begin{tabular}[c]{@{}l@{}}Didi dataset\\ GPS data, (private)\end{tabular} & \begin{tabular}[c]{@{}l@{}}Holdout (4-1 month)\end{tabular} \\
\rowcolor[HTML]{EFEFEF} 
Seq2Seq model~\cite{Karatzoglou2018ASL} & \begin{tabular}[c]{@{}l@{}}MIT Reality dataset\\ private dataset\end{tabular} & \begin{tabular}[c]{@{}l@{}}Holdout (70\%-30\% split)\end{tabular} \\
STF-RNN~\cite{AlMolegi2016STFRNNST} & Geolife & \begin{tabular}[c]{@{}l@{}}3-fold cross validation\end{tabular} \\
\rowcolor[HTML]{EFEFEF} 
ST-RNN~\cite{liu2016predicting} & \begin{tabular}[c]{@{}l@{}}GTD dataset\\ Gowalla dataset\end{tabular} & \begin{tabular}[c]{@{}l@{}}70\%-20\% split, 10\% validation\end{tabular} \\ \hline
\end{tabular}
}
\caption{Neural network approaches used to construct mobility prediction models, datasets used to corroborate model performance and the respective cross-validation strategies adopted.}
\label{tbl:rnn_variants}
\vspace{-15px}
\end{table}

In addition to the usage of different datasets for performance validation, we observe that the validation methodologies applied also differ widely from one another or bootstrapped using differing parameters. 
We categorize the currently used cross-validation techniques for assessing model predictability as: (1) random shuffling before holdout validation, (2) train-test split ratio ranging from 90\%-10\% to 50\%, (3) 3-fold to 30-fold cross-validation, and (4) weekday/month/year-based splitting. 
In the last case, model training is typically performed on the first four days of the week and tested on the remaining three.
We also found several occurrences where the validation methodology was not specified.
Selecting an appropriate validation approach to correctly assess the model performance on a given dataset is imperative to draw legitimate conclusions.
To the best of our knowledge, we did not find any argumentation to select a particular dataset or a particular validation strategy in the reviewed literature. 
In the next Section, we delve deeper in these shortcomings as we highlight that selecting arbitrary validation technique is detrimental to the incremental process of model improvement and provides misleading measures.

\begin{table}[t!]
\centering
\resizebox{0.47\textwidth}{!}{%
\begin{tabular}{lll}
\hline
\textbf{Approach} & \textbf{Dataset} & \textbf{Validation methodology} \\ \hline
\rowcolor[HTML]{EFEFEF} 
\begin{tabular}[c]{@{}l@{}}Template matching~\cite{Riedel2008RecognisingOS}\\ (SW-local alignment)\end{tabular} & Simulated data & \begin{tabular}[c]{@{}l@{}}Random selection, 30-fold validation\end{tabular} \\
\begin{tabular}[c]{@{}l@{}}Template matching~\cite{ossama2009similarity}\\ (Trajectory alignment)\end{tabular} & Simulated data & Not specified \\
\rowcolor[HTML]{EFEFEF} 
\begin{tabular}[c]{@{}l@{}}Data Mining~\cite{Monreale2009WhereNextAL}\\ (Decision tree)\end{tabular} & Car trajectories & \begin{tabular}[c]{@{}l@{}}Wednesday-train, Thursday-Test\end{tabular} \\
\begin{tabular}[c]{@{}l@{}}Data Mining~\cite{ma2018framework}\\ (Decision tree)\end{tabular} & \begin{tabular}[c]{@{}l@{}}Geolife dataset\\ Weibo (private)\end{tabular} & Not specified \\ \hline
\end{tabular}
}
\caption{Different data mining approaches used to model human-mobility, datasets used to corroborate the model's performance and the types of cross-validation strategies.}
\label{tbl:data_mining}
\vspace{-15px}
\end{table}

\section{Shortcomings}
\label{sec:shortcomings}

In this section, we experimentally investigate the ill-effects of the shortcomings described in Section~\ref{sec:review_summary}.
We categorize these shortcomings in two major domains: (1) inefficient accuracy vs. complexity trade-off arising from data-agnostic prediction model selection, (2) inconclusive model performance quantification due to adoption of inaccurate validation methodologies.
We also expose the systematic bias involved in the model assessment due to the selection of the dataset and validation methodology devoid of any heuristics. 

\subsection{Experimental Setup}

{\bfseries{Real world Mobility Datasets.}} We conduct the experiments by using three mobility datasets publicly available upon request.   
The PrivaMov dataset~\cite{privamov} was collected through GPS, WiFi and GSM sensors in the city of Lyon (France) and includes university students, staff and their family  members.
The Nokia mobile dataset~\cite{laurila2012mobile} (MDC) was collected in the Lake Geneva region of Switzerland and consists young individuals' trajectories, collected through GPS, WLAN, GSM and Bluetooth sensors.
The GeoLife dataset~\cite{zheng2010geolife} was collected in Beijing (China) and contains trajectories recorded through GPS loggers and GPS-phones.\newline
{\bfseries{Mobility Prediction}} can be defined as forecasting the transitions between places, after eliminating all self-transitions.
A preliminary step in achieving this consists of transforming the raw GPS locations into a sequences of points of interest (POIs).
A POI is defined as any location where an individual visits with an intentional purpose with a perceived priority for e.g., home/work place, gym, train station etc. 
We use a POI extraction technique that is independent of {\it{a priori}} assumptions regarding the data and individual mobility behaviors~\cite{Kulkarni:2017:EHW:3139958.3140002}. 
We then convert the raw GPS trajectory of a user $u$, $T_u = \langle(lat_1,lon_1,t_1), (lat_2,lon_2,t_2)...(lat_n,lon_n,t_n) \rangle$, where $lat_i, lon_i$ are the latitude and longitude coordinates respectively and $t_i$ is the timestamp such that $t_{i+1} > t_i$ into a sequence of temporally ordered points of interest, $s(t) = \langle (poi_1,t_1), (poi_2 t_2)...\rangle$, where $poi_i$ is the point of interest at index $i$.
The mobility prediction task is thus formulated as: given a sequence $s(t)$ up to a timestamp $n$, predict the next POI at timestamp $n+1$.\newline
{\bfseries{Prediction Accuracy}} is estimated by averaging the accuracy across all the individuals present in that dataset.
The individual prediction accuracy is computed by measuring the proportion of accurate predictions over all days of that individual (users who were not active on a day are excluded in the prediction).
The accuracy of a model can thus be formalized by Equation~\ref{eq:acc}.

\begin{equation}
\pi_{acc} = \frac{\sum_{t=1}^{T}\mathds{1}_{poi_t=poi_{t}^*}}{T},
\label{eq:acc}
\end{equation}

where $poi_t$ is the true next POI of an individual at time $t$, $poi_{t}^*$ is the predicted next point of interest and $T$ is the total number of prediction time-steps. 
The data is split into 10 windows consisting of 10\% training set and the subsequent 10\% as test set as performed in~\cite{Song2010LimitsOP}. 
The training is performed in a cumulative manner such that the previous training instance is not lost (more details in section~\ref{sec:data_driven_approach}).

\begin{figure}[t!]
    \centering
    \includegraphics[width=0.47\textwidth]{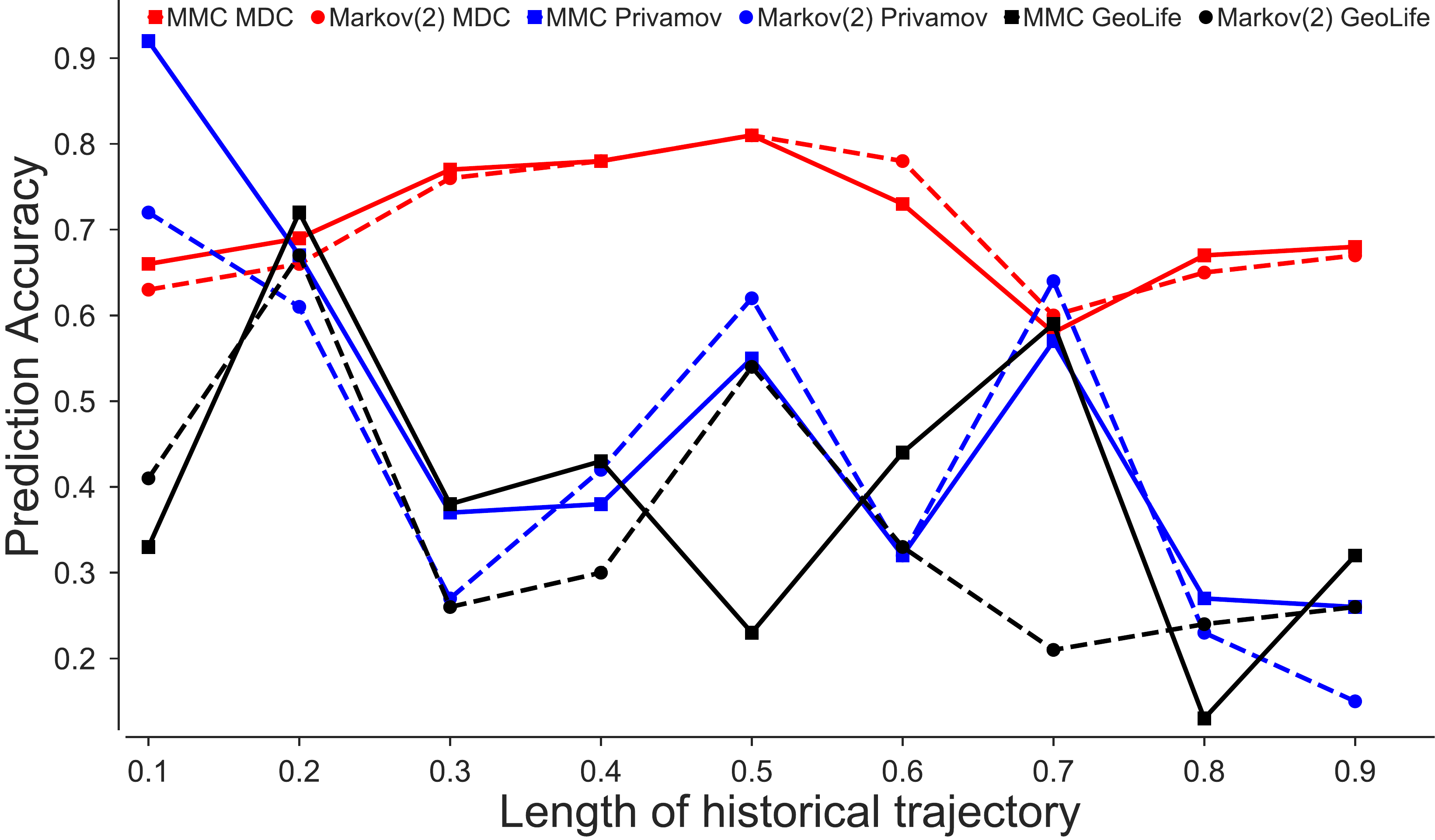}
    \caption{Prediction accuracy of Markov models variants on the three datasets. The horizontal axis signifies the proportion of trajectory length considered for the train-test split and vertical axis signifies the precision of the prediction model.}
    \label{fig:acc_comparison_mkv}
    \vspace{-10px}
\end{figure}

\subsection{Data-agnostic Model Selection}

In order to understand mobility dynamics at a high-granularity and to facilitate integration of these findings in mobility prediction models, it is paramount to evaluate distinct models of the same dataset.
Additionally, dataset selection for performance quantification without analyzing the dataset characteristics can mask critical aspects of the model behaviors. 
In this section, we focus on the following two aspects: (1) generalizability of model performance formulated using certain dataset(s), and (2) effects of selecting a model while ignoring dataset attributes on performance vs. complexity trade-off.

In order to analyze the generalizability of the model performance, we apply the same prediction model on mobility datasets differing widely with respect to some key characteristics.
We provide more details regarding these characteristics and quantify them in Section~\ref{sec:data_driven_approach} (see Table~\ref{tbl:dataset_characteristics}). 
We also compare the accuracy of different prediction models on the same dataset using a fixed cross-validation methodology.
Instead of presenting the average accuracy over the test-set, we adopt the canonical approach to perform sequential data cross-validation by rolling through the dataset (more details in Section~\ref{sec:data_driven_approach}).  
To facilitate easier result comprehension, we split the analysis first based on the usage of Markov models and neural networks.

\begin{figure}[t!]
    \centering
    \includegraphics[width=0.48\textwidth]{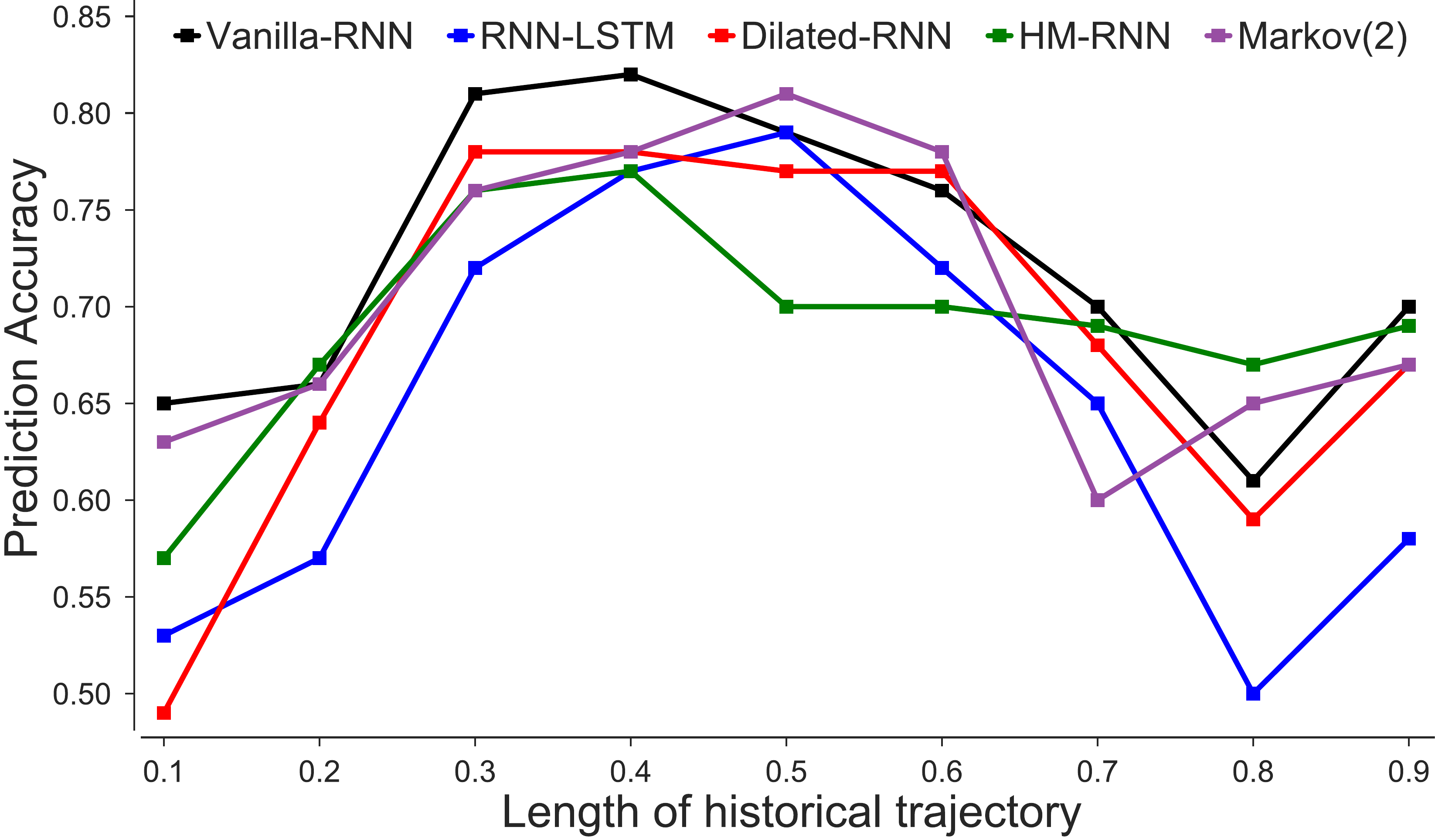}
    \caption{Prediction accuracy of RNN variants variants on the MDC dataset. The horizontal axis signifies the proportion of trajectory length considered for the train-test split and vertical axis signifies the precision of the prediction model.}
    \label{fig:mdc_comparison}
    \vspace{-15px}
\end{figure}

In Figure~\ref{fig:acc_comparison_mkv}, we compare the prediction accuracy of two approaches specified in Table~\ref{tbl:markov_model_variants}: (1) mobility Markov chain (MMC), and (2) second-order Markov model (Markov(2)).
We observe that the average accuracy and the variation trend of the Markov models differ by a large extent over the three datasets. 
Interestingly, the accuracy variation across the trajectory length is substantial for GeoLife and PrivaMov datasets as opposed to the MDC dataset. 
These accuracy variations stem from the fluctuation in dependencies between the POIs in a given dataset.
In Figure~\ref{fig:mdc_comparison}, we present the accuracy comparison of different RNN variants on the MDC dataset.
Although Vanilla-RNN and Markov-model (order-2) have lower computational complexity and lack long term memorization capability, they provide higher average accuracies as compared to RNN-variants.
The lower accuracies of RNN-variants stem from overfitting on the training set which results in dropping accuracy on the validation set.
We performed similar experiments on the PrivaMov and GeoLife datasets and observed that RNN-LSTM and HM-RNN provide the highest average accuracy on each respectively.
From these experiments, we emphasize that the performance of the same prediction approach can differ widely across datasets. 
Moreover, a prediction model that performs poorly on one dataset can provide sufficiently favorable results on another.  
As a result, it is clear that conclusions regarding algorithmic performance cannot be justified without defining the dataset characteristics.

Markov models differ in their capacity to incorporate $n$-previous locations and their temporal representation. 
The RNN extensions on the other hand differ in their capacity to manipulate the internal memory and propagate gradients along the network. 
This is due to the difference between the gating mechanisms employed, regularization techniques and the connections within the individual neurons and the hidden layers.
For instance, Dilated-RNNs~\cite{Chang2017DilatedRN} account for short and long-range correlations present in a sequence depending on the configuration of the skip-connections in the network. 
On the other hand, Hierarchical-Multiscale RNNs (HM-RNNs)~\cite{chung2016hierarchical} captures the latent hierarchical structure in the sequence by encoding the temporal dependencies with different timescales and thus being effective in representing the dependencies lying at different levels. 
This choice of the RNN variants was based on the ability of each to address distinct performance issues of the vanilla-RNN while modeling complex dependencies present in the dataset.  

\begin{table}[h!]
\centering
\resizebox{0.47\textwidth}{!}{%
\begin{tabular}{lll}
\hline
\textbf{Extension} & \textbf{Architecture} & \textbf{Features} \\ \hline
Vanilla-RNN~\cite{grossberg2013recurrent} & \begin{tabular}[c]{@{}l@{}}$\bullet$ no gating mechanism\\ $\bullet$ recurrent connections\end{tabular} & \begin{tabular}[c]{@{}l@{}}$\bullet$ faster,stable training\\ $\bullet$ simple architecture\end{tabular} \\ \hline
RNN-LSTM~\cite{Hochreiter1997LongSM} & \begin{tabular}[c]{@{}l@{}}$\bullet$ Vanilla-RNN connections\\ $\bullet$ cell gating mechanism\end{tabular} & \begin{tabular}[c]{@{}l@{}}$\bullet$ active self-connecting loops\\ $\bullet$ prevents memory degradation\end{tabular} \\ \hline
Dilated-RNN~\cite{Chang2017DilatedRN} & \begin{tabular}[c]{@{}l@{}}$\bullet$ LSTM cell structure\\ $\bullet$ dilated skip connections\end{tabular} & \begin{tabular}[c]{@{}l@{}}$\bullet$ parallelized computation\\ $\bullet$ long-term memorization\end{tabular} \\ \hline
HM RNN~\cite{chung2016hierarchical} & \begin{tabular}[c]{@{}l@{}}$\bullet$ variable dimensionality\\ $\bullet$ long credit assignments\end{tabular} & \begin{tabular}[c]{@{}l@{}}$\bullet$ hierarchical-temporal representation \\ $\bullet$ ovel update mechanism\end{tabular} \\ \hline
\end{tabular}%
}
\caption{Recurrent Neural Network variants with their respective architectural differences and features.}
\label{tbl:rnn_archs_procon}
\vspace{-5px}
\end{table}

We present a summary of the major architectural differences and the features associated with each extension selected in this work to carry of experiments in Table~\ref{tbl:rnn_archs_procon}.
We use the standard implementations of the predictive algorithms as described in their respective papers, i.e., the same architecture (\#neurons, \#hidden layers, vertical depth) and the same hyperparameters (learning rate, unroll steps, activation function, optimizer and dropout rate).  
Mobility Markov chains~\cite{gambs2012next} and Markov models order(2)~\cite{si2010mobility} are implemented using the standard libraries.\footnote{hmmlearn: \url{https://hmmlearn.readthedocs.io}}
Vanilla-RNN~\cite{grossberg2013recurrent}, RNN-LSTM~\cite{Hochreiter1997LongSM} and dilated-RNN~\cite{Chang2017DilatedRN} are based on predicting the next character (language modeling) in the text, whereas HM-RNN~\cite{chung2016hierarchical} models the prediction task as multivariate classification. 
For dilated-RNN~\cite{Chang2017DilatedRN} we use the dilations of 1, 2, 4, 8, 16, 32 and 64 and provided the results for dilation 32 after which we observe a drop in the accuracy.

We repeat the experiment shown in Figure~\ref{fig:mdc_comparison} with the PrivaMov and GeoLife datasets.
We observe that on the PrivaMov dataset, RNN-LSTM and dilated-RNNs provide a comparable accuracy and outperform the other models.
However, HM-RNN provides the best accuracy on the GeoLife dataset. 
This leads us to question the generalizability of the performance results specified in the current studies.
We thus argue that data agnostic model selection often results in an ineffective trade-off which necessitates a mobility data attribute-driven modeling approach.

\subsection{Flawed Validation Methodology}

\begin{figure}[t!]
    \centering
    \includegraphics[width=0.47\textwidth]{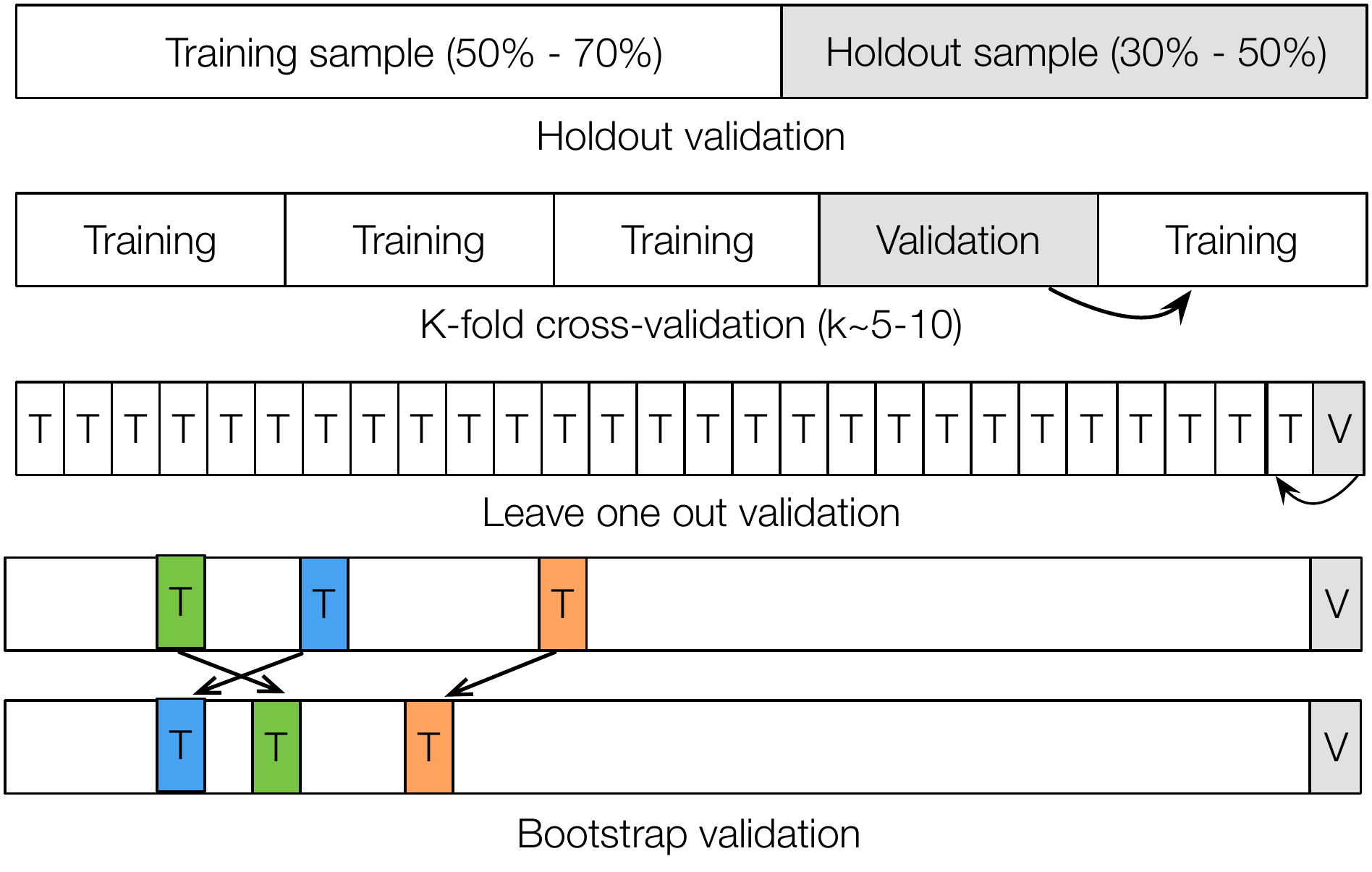}
    \caption{Different methodologies of cross-validation. The figure shows the proportion and position of train-set (white) and the validation set (grey).}
    \label{fig:cross_validation}
    \vspace{-15px}
\end{figure}

In this Section, we focus on the second shortcoming that relates to the usage of inaccurate validation strategies by the existing studies and we present the misleading effects of their adoption. 
Table~\ref{tbl:markov_model_variants},~\ref{tbl:rnn_variants},~\ref{tbl:data_mining} show distinct validation strategies used to validate the proposed prediction models. 
These strategies can be categorized as follows: (1) holdout cross-validation, (2) leave one out cross-validation, (3) k-fold cross-validation, and (4) bootstrapping. 
Figure~\ref{fig:cross_validation} presents the overview of these approaches. 
Holdout is the most common cross-validation techniques used where the data is typically split 70\%-30\% for training and testing; the model accuracy is reported on the test-set.
In the case of k-fold cross-validation, the dataset is first split into $k$ equal partitions and the training is performed on all but one, which is the validation set.
This set is replaced $k$ times with other partitions of the dataset to perform validation, the average accuracy on the $k$ sets is then reported. 
The leave one out cross-validation technique is similar to k-fold, where a single data-point is treated as a sub-sample. 
The bootstrapping technique performs the train-test split after resampling the data points with replacement from the original dataset for every validation iteration.

Although the above validation approaches are suitable for accessing model performance on the dataset devoid of any temporal component, they cannot be applied for mobility trajectories as it is a sequential form of data. 
In order to validate prediction models based on sequential data, no future observations can be used to train the model. 
Furthermore, the data cannot be randomly split into train-test groups due to the temporal dimension of the geolocation observations. 
Instead, the data must be split with respect to the temporal order in which the values were observed.
These approaches are highly susceptible to selection bias if the distribution of the data in the train-test set is not identical.  
We further discuss the implications of these validation techniques and propose adoption of a standardized technique in Section~\ref{sec:data_driven_approach}.

\begin{table}[t!]
\centering
\resizebox{0.45\textwidth}{!}{%
\begin{tabular}{llll|lll}
\hline
Dataset & \multicolumn{3}{l|}{Holdout cross-validation} & \multicolumn{3}{l}{K-fold cross-validation} \\ \hline
 & 80-20 & 70-30 & 60-40 & 3-fold & 5-fold & 10-fold \\ \cline{2-7} 
MDC & 0.78 & 0.81 & 0.66 & 0.63 & 0.72 & 0.65 \\
\rowcolor[HTML]{EFEFEF} 
PrivaMov & 0.63 & 0.65 & 0.45 & 0.68 & 0.57 & 0.52 \\
GeoLife & 0.83 & 0.65 & 0.63 & 0.75 & 0.70 & 0.61 \\ \hline
\end{tabular}%
}
\caption{Prediction accuracies derived by using different splits for holdout validation and different values of $k$ for k-fold cross-validation.}
\label{tbl:cross_validation_comparison}
\vspace{-15px}
\end{table}

In order to highlight the misleading nature of the above validation approaches in the context of application to mobility modeling, we apply the holdout and k-fold cross-validation to access the performance of a Vanilla-RNN model (see Table~\ref{tbl:cross_validation_comparison}).
We select Vanilla-RNN due to discernibility compared to other RNN variants. 
We observe that different train-test split ratios result in different prediction accuracies in case of holdout validation for all the three datasets under consideration. 
A similar behavior is observed in case of k-fold cross-validation for distinct values of $k$. 
Thus, it is clear that the accuracy results computed by these validation measures are neither a conclusive evidence of model performance, nor do they provide a comparative measure to analyze performance with respect to another model. 
Based on the above, we argue that application of flawed validation methodology is detrimental to the advancement of human-mobility modeling.

\section{Mobility-Modeling Framework}
\label{sec:data_driven_approach}

In this section, we address the shortcomings in the reviewed studies described in Section~\ref{sec:shortcomings}.
We first present the data-driven mobility modeling framework and introduce mobility meta-attributes selected through experimentation based on rigorous statistical tests. 
We also present the block-rolling validation methodology to correctly assess the performance of the prediction model.

\begin{figure}[t!]
    \centering
    \includegraphics[width=0.4\textwidth]{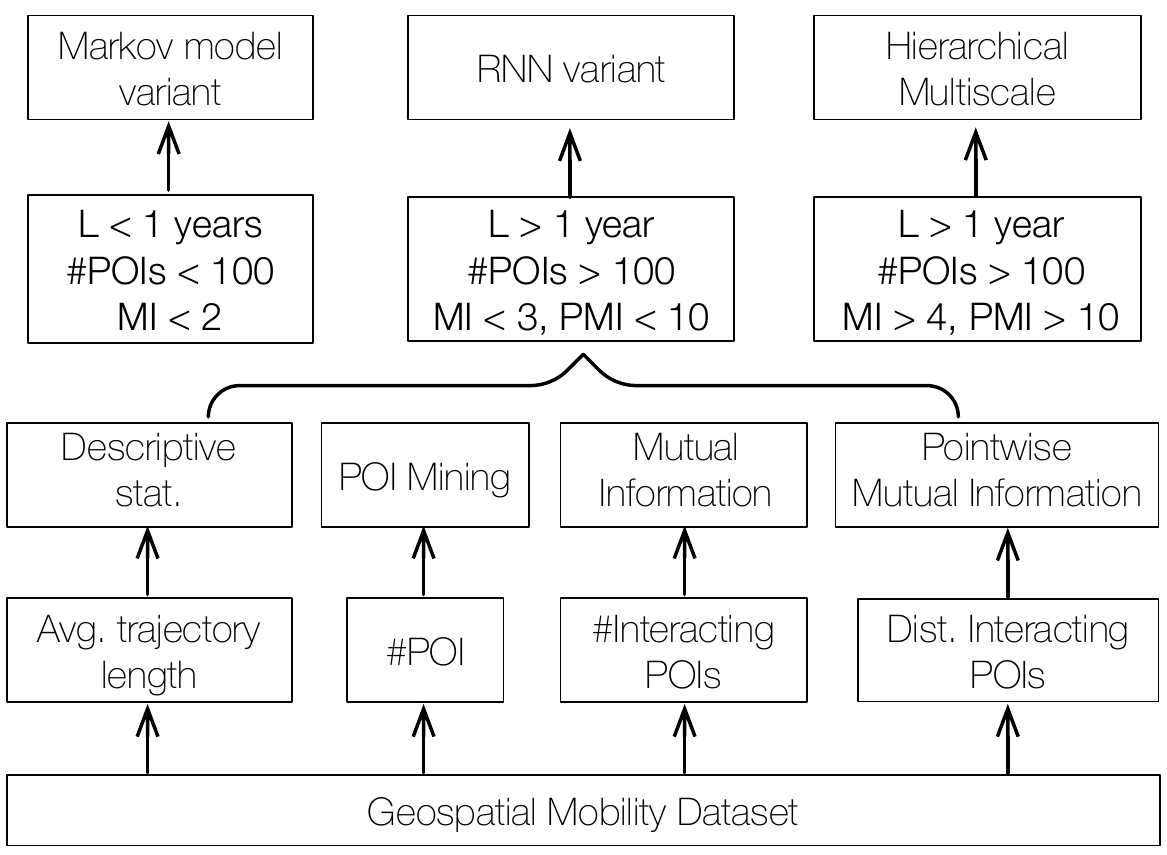} 
    \caption{Data-drive mobility model selection framework.}
    \label{fig:data_driven}
    \vspace{-10px}
\end{figure}

Figure~\ref{fig:data_driven} presents our approach for data-driven selection of the modeling technique. 
We identify the following four meta-attributes to quantity the characteristics of a mobility dataset: (1) Average trajectory length, (2) \#POIs, (3) \# POI interactions, and (4) POI interaction distance.
Figure~\ref{fig:data_driven} also shows the experimentally determined attribute thresholds to select a particular approach.

\subsection{Meta-Attribute Selection}

Mobility meta-attributes aid in characterizing a dataset to predict performance of forecasting algorithms and subsequently in selecting an appropriate modeling algorithm.
We use statistical and information theoretic measures to estimate these meta-attributes and present the general descriptive statistics of the three considered mobility datasets in Table~\ref{tbl:dataset_characteristics}.
The temporal mobility entropy is estimated using the approach specified by Song et al. in their seminal work regarding quantification of predictability limits~\cite{Song2010LimitsOP} based on Lempel-Ziv data compression~\cite{ziv1978compression}.
We  refer the readers to~\cite{kulkarni2019examining} for further reading regarding computation of mobility entropy and subsequently determining the predictability.

\begin{table}[h!]
\centering
\resizebox{0.48\textwidth}{!}{%
\begin{tabular}{cccccccc}
\hline
\textbf{Dataset} & \textbf{\#users} & \textbf{\#months} & \textbf{traj. length} & \textbf{POIs} & \textbf{Granularity} & \textbf{entropy} & \textbf{predictability} \\ \hline
PrivaMov & 100 & 15 & 1560000 & 2651 & \begin{tabular}[c]{@{}c@{}}246 meters\\ 24 seconds\end{tabular} & 7.63 & 0.7646 \\
\rowcolor[HTML]{EFEFEF} 
MDC & 191 & 24 & 685510 & 2087 & \begin{tabular}[c]{@{}c@{}}1874 meters\\ 1304 seconds\end{tabular} & 6.08 & 0.8323 \\
GeoLife & 182 & 36 & 8227800 & 3892 & \begin{tabular}[c]{@{}c@{}}7.5 meters\\ 5 seconds\end{tabular} & 9.77 & 0.6219 \\ \hline
\end{tabular}%
}
\caption{Descriptive statistics with entropy and predictability.}
\label{tbl:dataset_characteristics}
\vspace{-10px}
\end{table}

\begin{figure}[t!]
    \centering
    \includegraphics[width=0.4\textwidth]{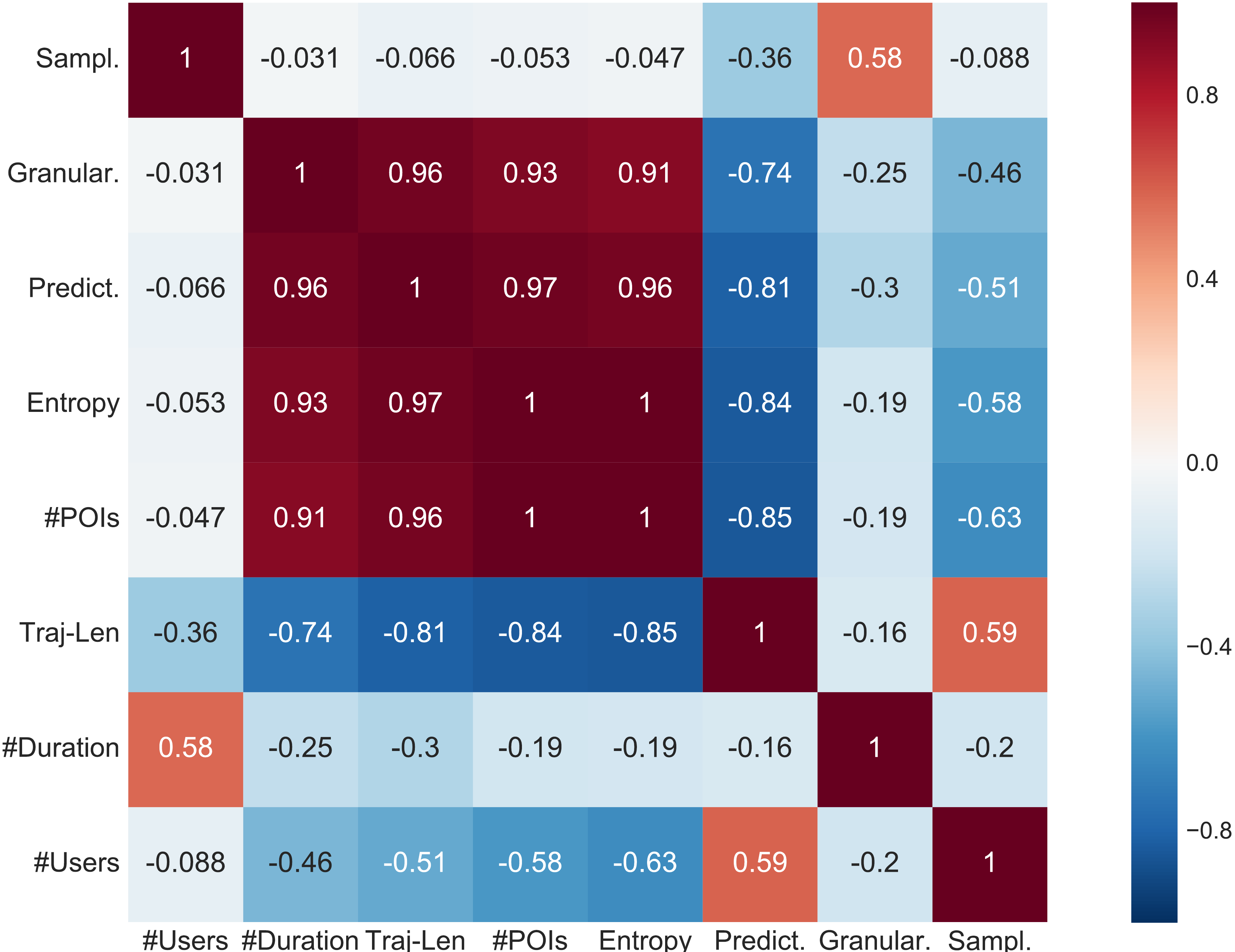}
    \caption{Correlation matrix for all the descriptive statics variables, entropy and predictability}
    \label{fig:corr_mat}
    \vspace{-10px}
\end{figure}

In order to analyze the correlations between the characteristics presented in Table~\ref{tbl:dataset_characteristics}, we first segment the user trajectories from all the considered datasets into substrings of lengths ranging from 1 month to 60 months.
The partitioned and aggregated datasets are then used to compute the descriptive statistics presented in Table~\ref{tbl:dataset_characteristics}. 
The above step is performed in order to generate a large amount of mobility data from to achieve higher confidence in the statistical tests. 
We first analyze variables in Table~\ref{tbl:dataset_characteristics} to find significant correlations with predictability (see the correlation matrix in Figure~\ref{fig:corr_mat}). 
We find a significant correlation between the duration (length) of the user trajectory and mobility predictability. 
We find that trajectories collected for longer durations incorporate varying mobility behaviors, quantified in terms of periodicity variation with the POIs. 
However, this effect is moderated by the number of POIs which determine the predictability.  
Therefore, trajectories spanning longer duration along with growing number of POIs shows a significant positive correlation with the mobility entropy and hence predictability. 
The t-test avail a p-value of $0.0002$ for predictability and trajectory length and $0.000045$ for predictability and number of POIs. 
In general, if we consider two mobility datasets $D_1$ and $D_2$, collected for time durations $t_1$ and $t_2$, where $t_1 > t_2$, the number of POIs in $D_1 > D_2$ 
This would result in the the predictability of $D_1 < D_2$ due to higher entropy of $D_1$ because of the varying periodicities in larger number of POIs.

In order to understand the influence of the POIs and the trajectory length in more detail, we estimate the dataset structure to analyze the relationship between groups of POIs (n-grams) and their periodicities. 
To derive the structure, we arrange the dataset in terms of stacked sequences belonging to each user, where each sequence contains time ordered POIs, henceforth termed as {\it{symbols}}.
We base this approach on Yan et al.~\cite{yan2013diversity} work on aggregating individual mobility patterns to analyze aggregate scaling properties. 
Due to space restrictions, we only present the dataset structure of the PrivaMov dataset in Figure~\ref{fig:comparison_fv}. 
We upload the high resolution dataset structure images on our github page of this project~\footnote{Github Page: \url{https://bit.ly/2HRZGk5}}.
The dataset structure is with regards to symbol repetition statistics adapted from the file viewer utility contributed by Matt Mahoney~\footnote{File Viewer (fv): http://mattmahoney.net/dc/textdata.html}.
We observe the distinct repetitive structure in each dataset at different levels and for different symbol-lengths. 
In the case of MDC dataset, we see symbol repetitions of length 1 occurring at a distance ranging between 1 to 10, and a few symbol matches of length 4 separated by at least $10^1$ symbols. 
In the case of PrivaMov and GeoLife dataset however, the blue bands at the top shows that symbol matches of length 8 often separated by a distance of $10^5$. 
The green band shows matches of length 4 commonly separated by $10^3$ to $10^4$ symbols, whereas the red bands show that the matches of length 2 are separated by about 10 to 500 symbols. 
The grey band shows that single symbol matches are usually separated by a distance of 1, 3 or 10.

Based on the dataset structure analysis, we see that there are longer dependencies spanning larger symbol lengths in case of GeoLife dataset as compared to PrivaMov dataset, whereas the MDC dataset mostly contain short-term dependencies.
We therefore propose to leverage long-distance dependencies present in a dataset as another means to quantify the dataset characteristics.
Therefore in addition to the average trajectory length and the number of unique POIs, we include LDDs as a meta-attribute.

\subsection{Long-distance Dependencies}

A long-distance dependency describes a contingency or interaction between two or more elements in a sequence that are separated by an arbitrary number of positions.
More formally, LDDs are related to the rate of decay of statistical dependence of two symbols with increasing time interval or spatial distance between them.
LDDs are commonly observed in natural languages, for instance in English, there is a requirement for the subjects and verbs to agree, i.e., words bear relations and interact with other words in every sentence.
Such a relation valuate one item with respect to the other within a certain search space. 
We extend this concept to human mobility where the POIs can be viewed as symbols in a natural language. 
Thus, mobility trajectory might display different degree of LDD depending on an individuals behavior, thus making them challenging to model computationally.

\begin{figure}[t!]
    \centering
    \includegraphics[width=0.48\textwidth]{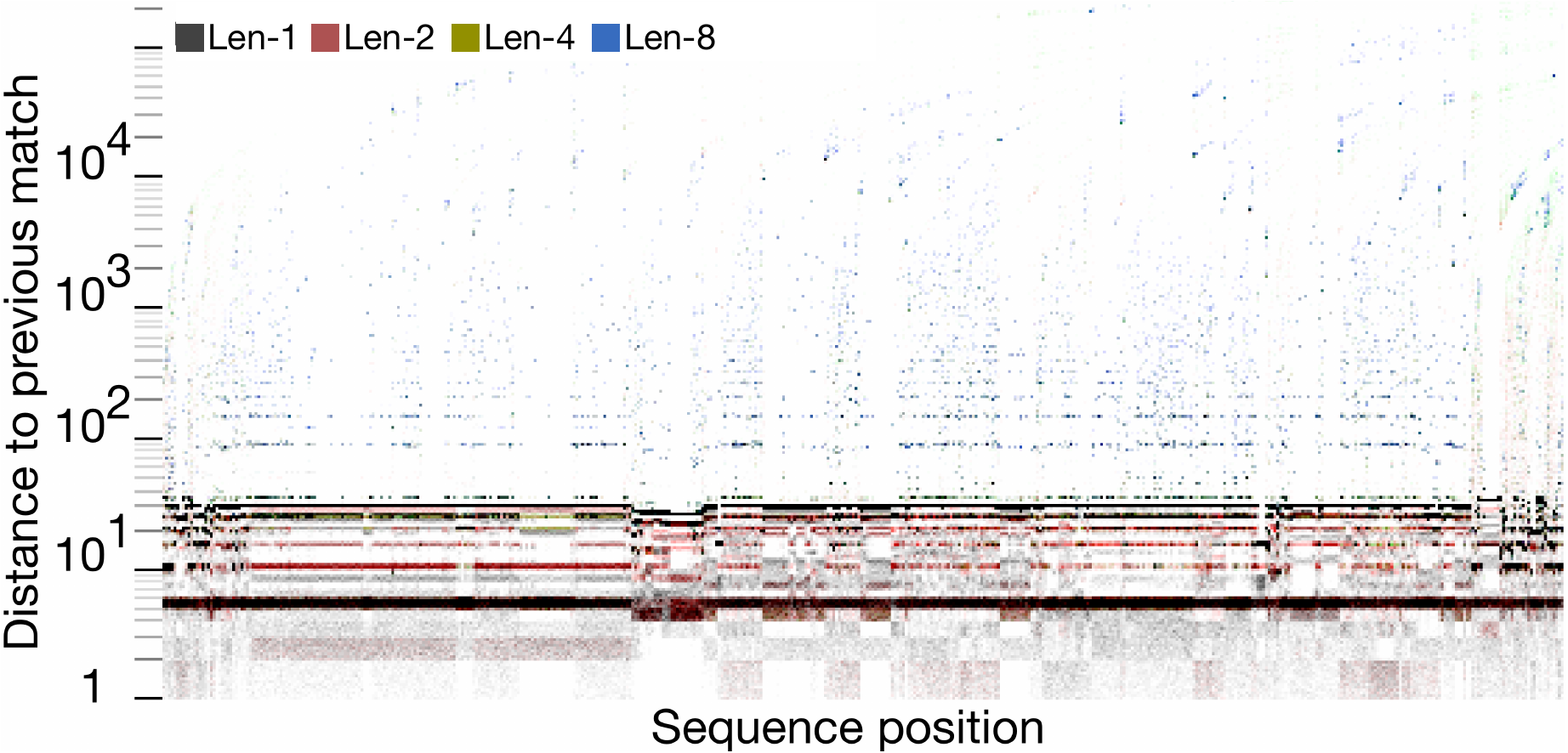}
    \caption{The figure shows the distribution of symbol (POI) matches of length (1) black, 2 (red), 4 (green), and 8 (blue) in the PrivaMov dataset. The horizontal axis represents the position of the symbol sequence in the dataset, whereas the vertical axis shows the distance backwards to the previous match on a logarithmic scale.}
    \label{fig:comparison_fv}
    \vspace{-15px}
\end{figure}

{\bfseries{Mutual Information.}} Computing the mutual information of the data under consideration can be seen as a statical framework for discerning and quantifying the presence of LDDs.
Mutual information $I$ is a quantity that measures the relationship between two symbols and quantifies the measure of information communicated, on average by one symbol about another.
$I$ as a function of distance between the individual events indicates the distribution of large but rare events and identify the presence of memory in the sequence.
Mutual information between symbols $X,Y$ is given by Equation~\ref{eq:mi}.

\begin{equation}
\begin{aligned}
\label{eq:mi}
I(X;Y) &= \sum_{X,Y}p(X,Y)\log\frac{p(X,Y)}{p(X).p(Y)} \\
& = H(X) - H(X|Y) \\
& = H(Y) - H(Y|X) \\
& = H(X) + H(Y) - H(X,Y),
\end{aligned}
\end{equation}

where $p(X,Y)$ is the joint distribution of two random variables $X$ and $Y$, $p(X)$ and $p(Y)$ are the marginal distributions of $X$ and $Y$. 
$H(X,Y)$ is the joint entropy of two random variables, $X,Y$ distributed according to the $pmf$ $p(X,Y)$ and $H(X|Y)$ is the conditional entropy of $X$ given $Y$.
Mutual Information can thus be used to quantify the interactions between the POIs in the dataset.

{\bfseries{Pointwise Mutual Information.}} The strength of the interaction between individual symbols can then be estimated using a related concept known as pointwise mutual information (PMI).
Unlike $I$, which quantifies the average information communicated by one symbol in the sequence about another, PMI quantifies the actual probability of co-occurrence of events $p(X,Y)$ differing from the expectation.
It is computed on the basis of the probabilities of the individual events under the assumption of independence $p(X)p(Y)$ according to Equation~\ref{eq:pmi}.

\begin{equation}
PMI(X,Y) = log_2\frac{N.C(X,Y)}{C(X).C(Y)}
\label{eq:pmi}
\end{equation}

$PMI(X,Y)=0$ indicates that $X$ and $Y$ are statistically independent.
Here, $C(X)$ and $C(Y)$ is the total number of occurrences of $X$ and $Y$ respectively and $C(X,Y)$ is the co-occurrence of $(X,Y)$.

\begin{figure}[t!]
    \centering
    \includegraphics[width=0.47\textwidth]{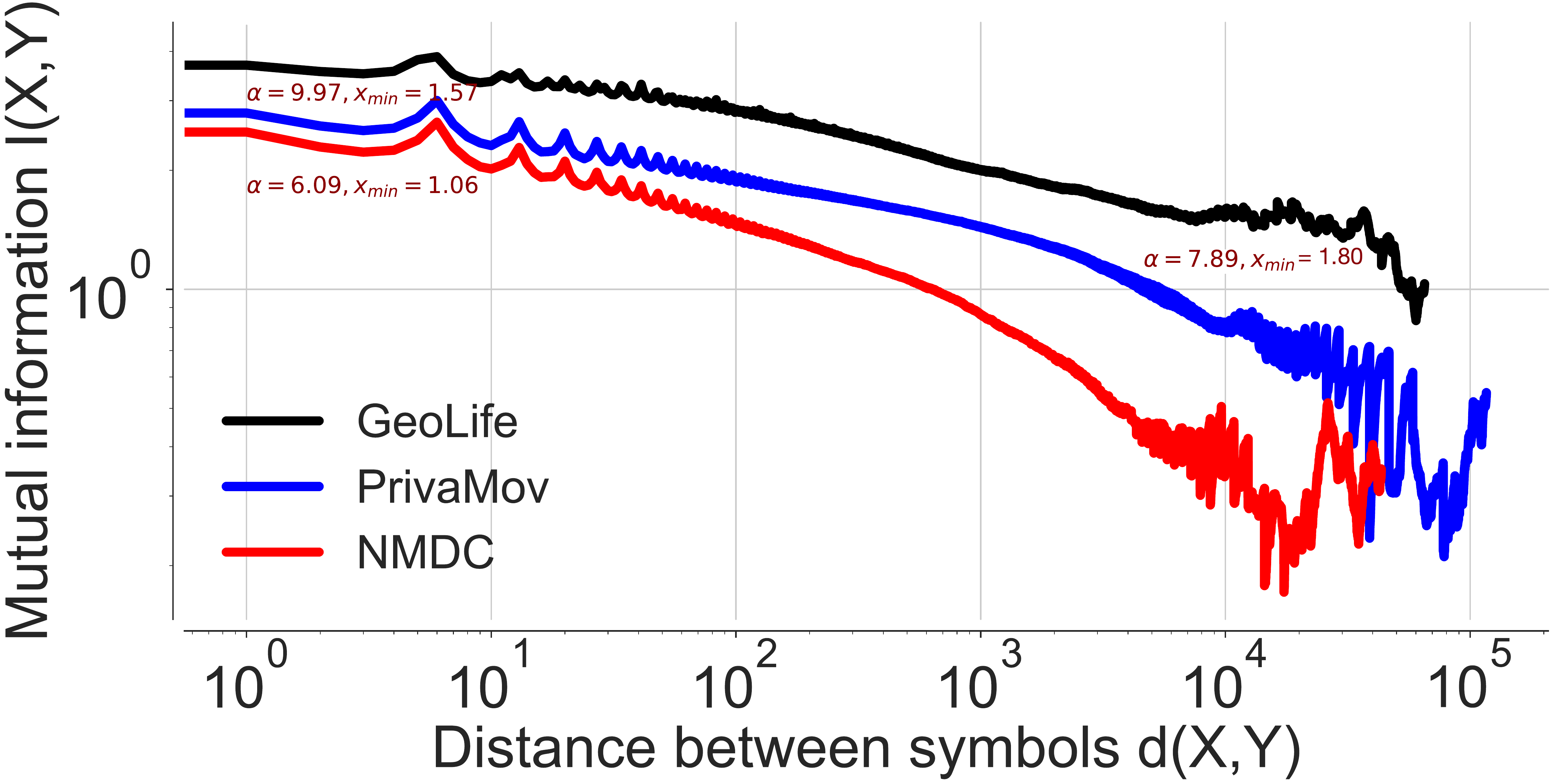}
    \caption{Mutual information decay in the three datasets. The vertical axis represents the bits per symbol as a function of separation $d(X,Y)=|i-j|$, where the symbols $X$ and $Y$ are located at positions $i$ and $j$ in the considered sequence.}
    \label{fig:mi_alldatasets}
    \vspace{-15px}
\end{figure}

Figure~\ref{fig:mi_alldatasets} shows the LDD characteristics of all the three datasets considered in this work.
All the measured curves for the three datasets are seen to decay roughly as power laws, and the value of exponent $\alpha$ indicates the extend of LDDs (power-law with cut-off for MDC dataset).
Recall from Figure~\ref{fig:comparison_fv}, where  we noted a trend in LDDs, which is corroborated by the mutual information analysis in Figure~\ref{fig:mi_alldatasets}.
We also observe the effect of LDDs on the prediction accuracy results presented in Figure~\ref{fig:acc_comparison_mkv} and Figure~\ref{fig:mdc_comparison}.
The MDC dataset provides higher accuracy as compared to the other two datasets and has a lower variation within the accuracies of different algorithms. 
This stems from the presence of short-distance dependencies in the individual trajectories present in the dataset (see Figure~\ref{fig:mi_alldatasets}).
Analyzing the mutual information trend also sheds light on the reasons pertaining to lower accuracies provided by RNN architectures, compared to Markov models at certain positions of trajectory-lengths.
One reason could be the tendency of RNN models to actively seek for long-range dependencies while overlooking the short-term dependencies. 
We validate this behavior in the case of dilated-RNN's, where an increase in dilations (to account for longer dependencies) results in lowering prediction accuracies.

\begin{figure*}[t!]
\centering
\subfloat[MDC Dataset]{\label{fig:mdc_validation} \includegraphics[scale=0.17]{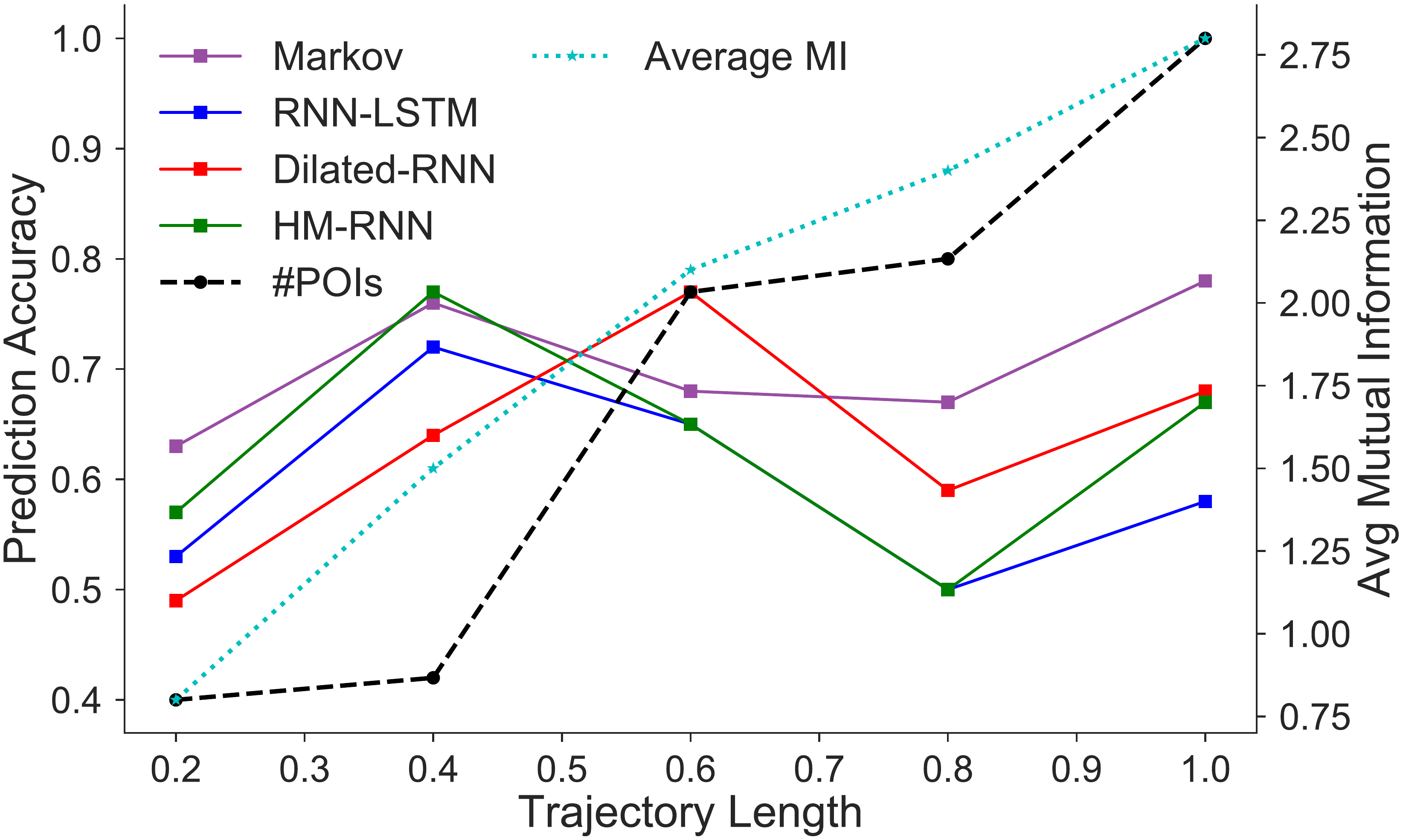}} 
\subfloat[PrivaMov Dataset]{\label{fig:privamov_validation} \includegraphics[scale=0.17]{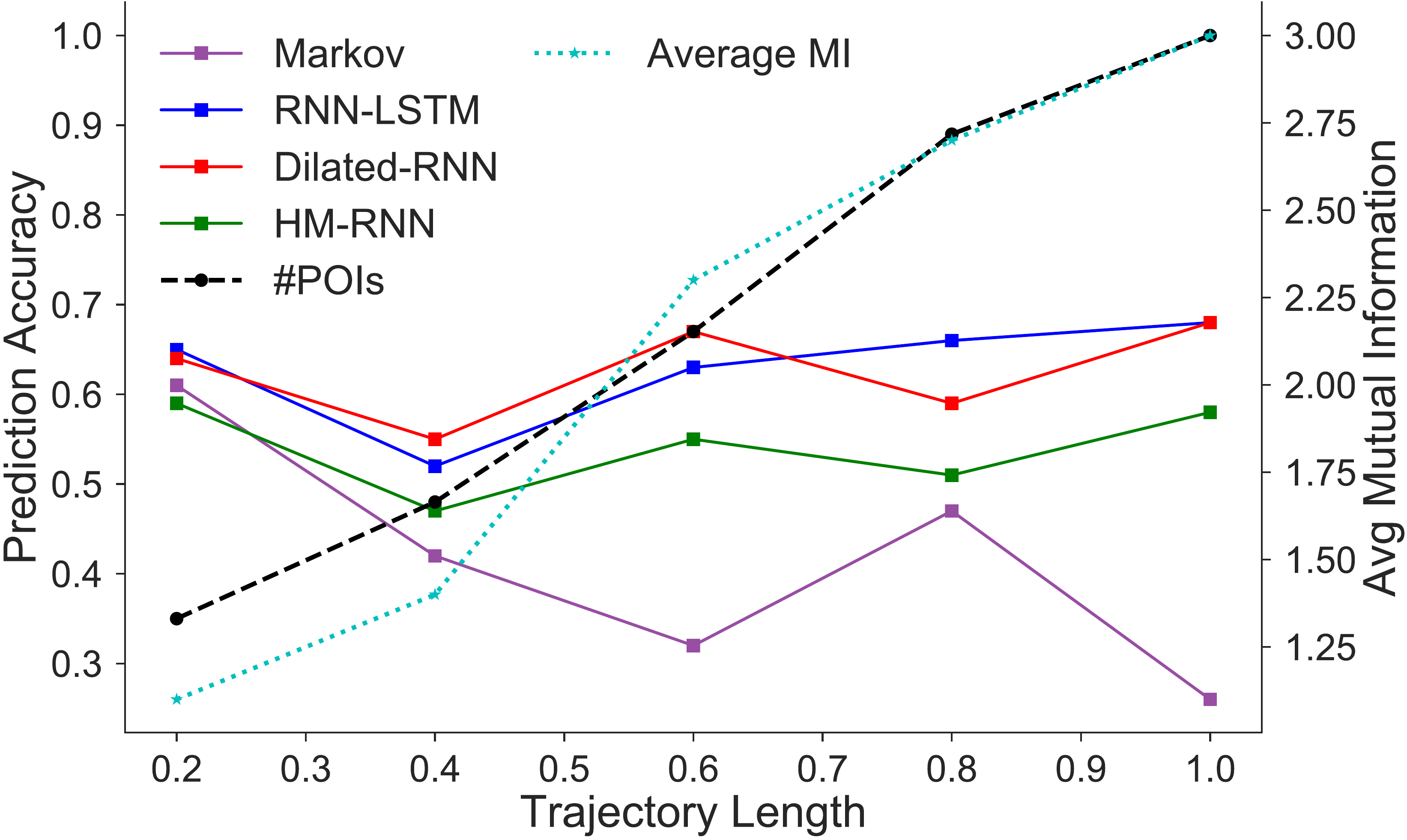}}
\subfloat[GeoLife dataset]{\label{fig:geolife_validation} \includegraphics[scale=0.17]{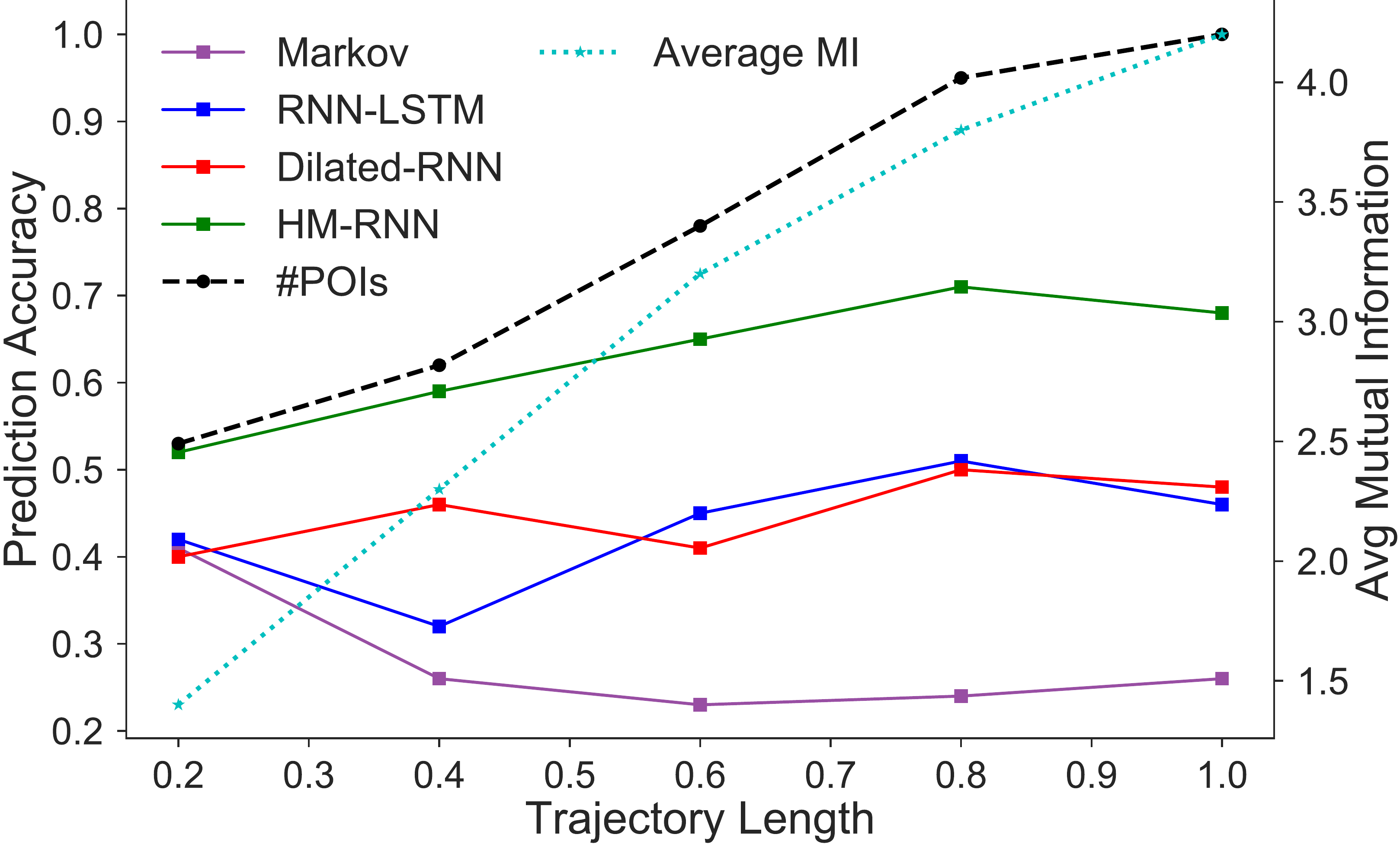}}
\caption{Experimental validation of the proposed framework by analyzing the prediction accuracy and its relationship with \#POIs and dependency depth (mutual information).} 
\label{fig:experimental_validation} 
\end{figure*}

\begin{figure*}[t!]
\centering
\subfloat[MDC Dataset]{\label{fig:mdc_cr} \includegraphics[scale=0.17]{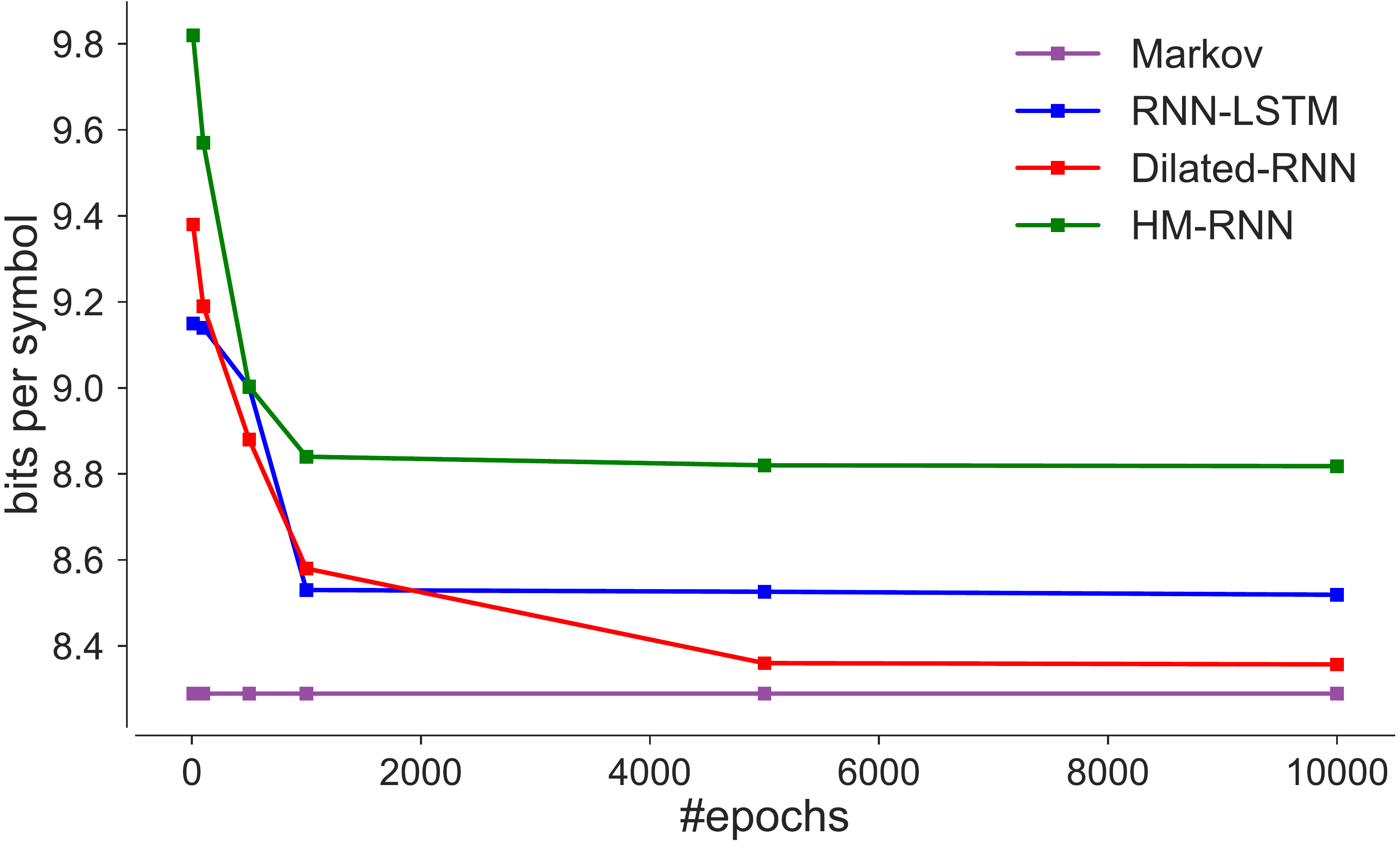}} 
\subfloat[PrivaMov Dataset]{\label{fig:privamov_cr} \includegraphics[scale=0.17]{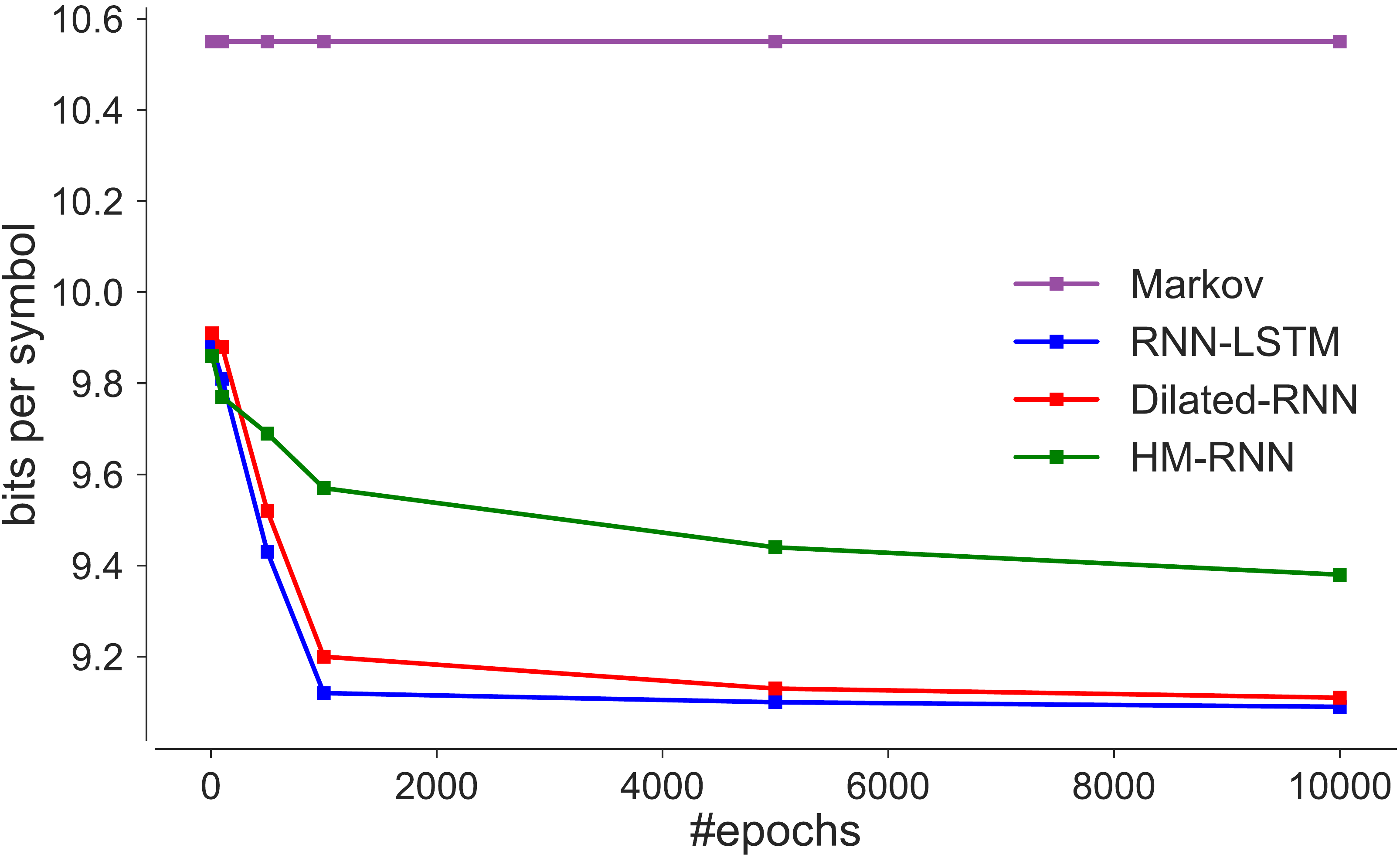}}
\subfloat[GeoLife dataset]{\label{fig:geolife_cr} \includegraphics[scale=0.17]{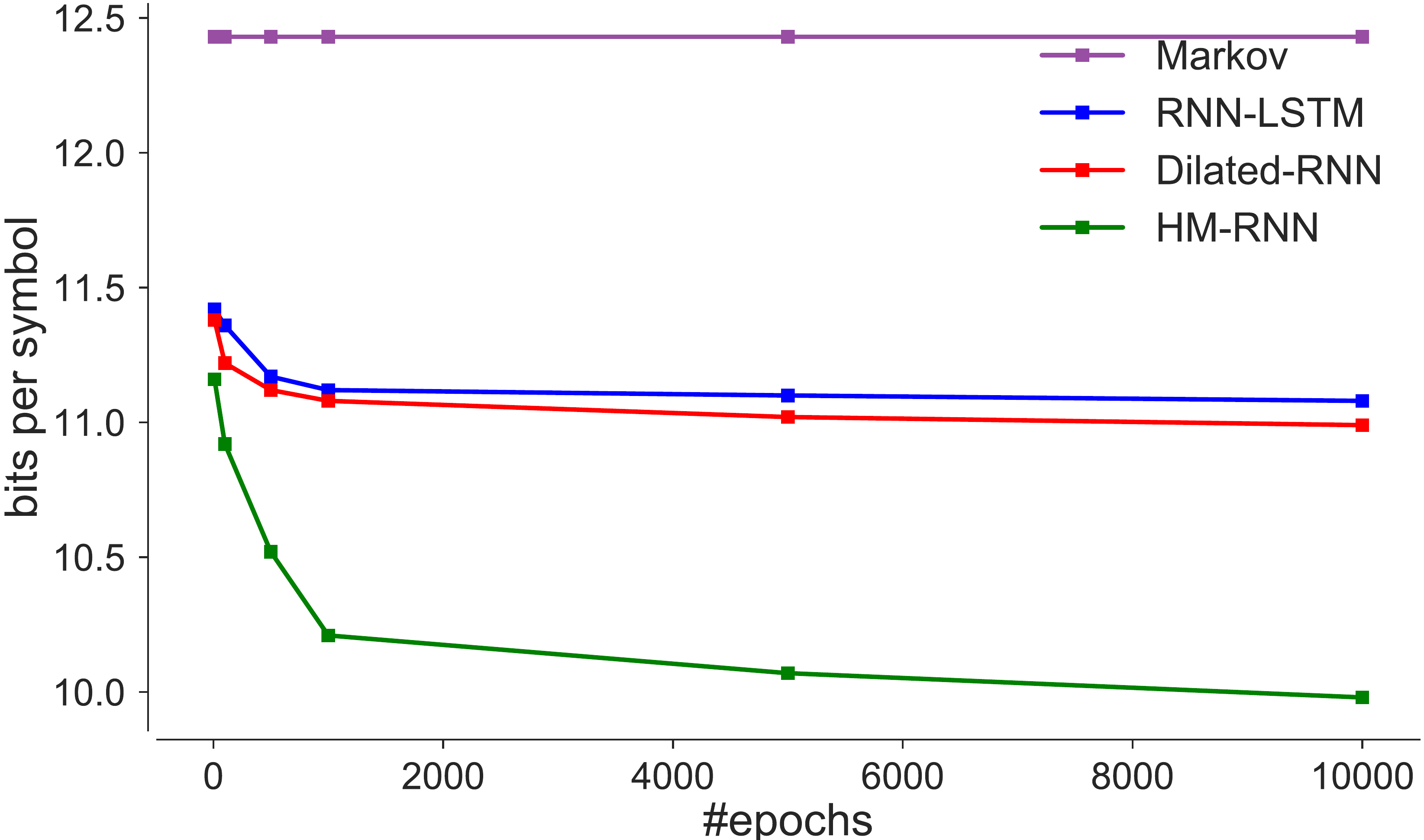}}
\caption{Compression ratio computed by each prediction model in terms of bits/symbol (POI).} 
\label{fig:compression_ratio} 
\end{figure*}

\subsection{Validation Methodology}

In order to assess to performance of mobility modeling techniques, we adopt the block-rolling validation strategy used to validate time-series prediction  models~\cite{bergmeir2012use}.
Sequential data such as mobility trajectories is subjected to autocorrelation~\cite{bergmeir2012use}, where the assumption made by the currently used validation approaches of i.i.d observations does not hold.
Therefore, techniques such as holdout and k-fold cross-validation cannot be applied.
For instance, 3-fold cross-validation applied over 3 time periods ignores the sequential nature of time, mixing up the past, present and future trajectory data points. 
Application of leave-one out or bootstrap is also not valid in this scenario as filtering out a data point does not remove the associated information due to the correlations with other observations. 

\begin{figure}[t!]
    \centering
    \includegraphics[width=0.4\textwidth]{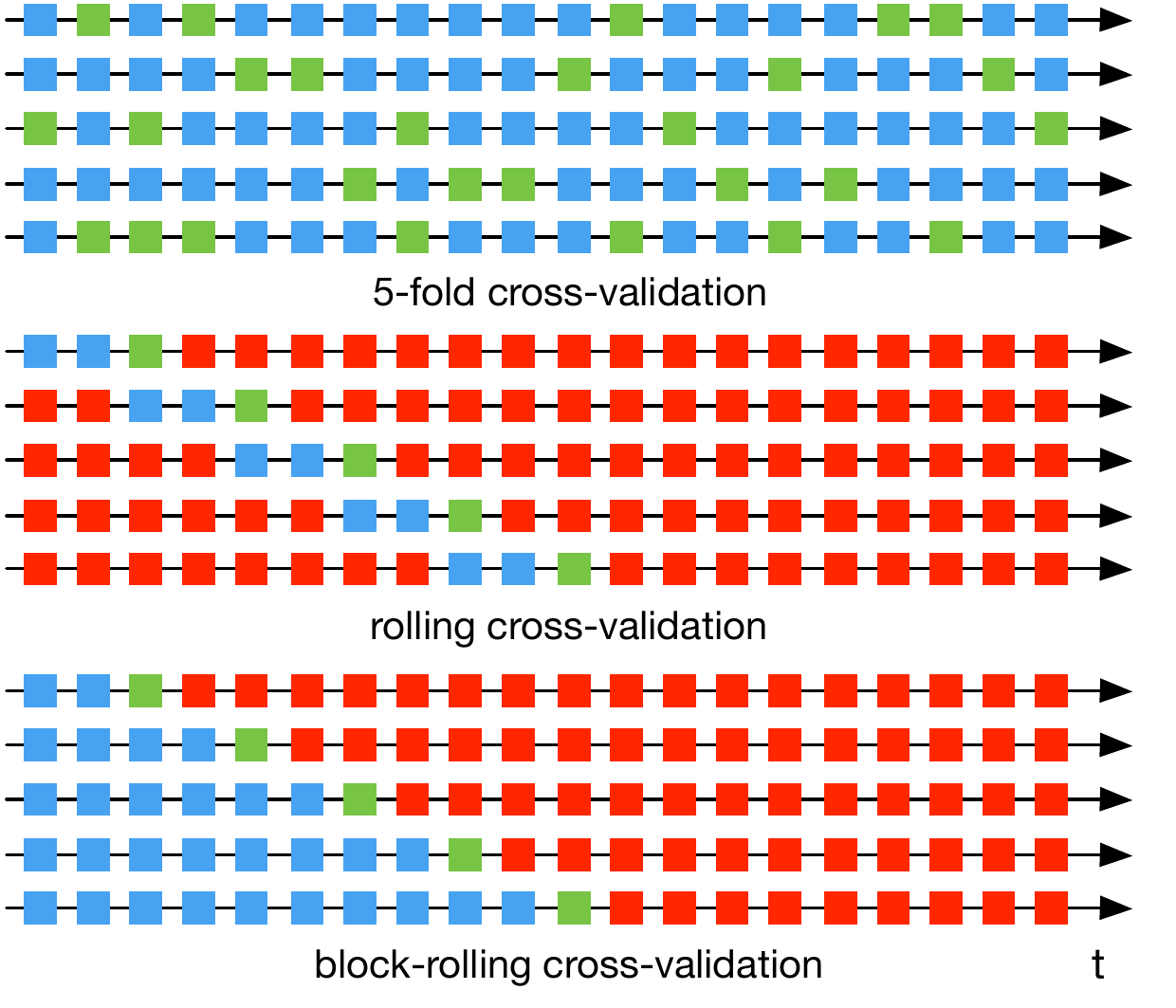}
    \caption{Comparison of the 5-fold cross validation, rolling cross-validation, and block-rolling cross-validation techniques. The blocks in blue indicate data-points seen by the training model, the green blocks indicate the validation data-points and the red blocks are not seen by the training model.}
    \label{fig:block_roll}
    \vspace{-15px}
\end{figure}

In the block-rolling validation strategy, the dataset is split into $k$equal size blocks.
The train set always consist of $p$ contiguous blocks and the validation is performed on the block $p+1$ as shown in Figure~\ref{fig:block_roll}.
In the case of rolling cross-validation strategy, the blocks might be partitioned to include POI pairs that change minimally in their visitation periodicity, but frequently in time.
The same partitions can include POI pairs that do not change in their periodicity over a long time periods, making this type of splitting misleading.  
The block-rolling technique incorporates these changes in the long-running variable and hence provides an unbiased validation after averaging over all the test sets.

\section{Evaluation}
\label{sec:results_discussion}

In this section, we experimentally assess the trade-offs involved in model selection, following out data-driven modeling framework. 
We first compute the prediction accuracies using the three classes of modeling techniques: (1) Markov model, (2) RNNs, and (3) HM-RNN.
We compare the block-rolling accuracies at each stage (every segment of trajectory length splits), with the number of POIs and the LDDs (mutual information) in Figure~\ref{fig:experimental_validation}. 
For each dataset, we run the model 5 times  and report the highest accuracy of the RNN and the HM-RNN models. 
We can see that Markov model (order-2) performs reasonably well on MDC dataset containing very short dependencies and lower number of POIs as compared to the other datasets.
However, when the LDDs/dependency depth exceeds 2, we find that the performance of Markov model drops very quickly and comparable with random guess's performance ($\approx 10\%$ variance).
On the other hand, the performance of RNN-LSTM and Dilated-RNNs drops after mutual information exceeds 3, but is substantially better than random guessing. 
When the mutual information is between 2 and 3, LSTMs and Dilated RNN perform similarly, however Dilated-RNNs demonstrate an unstable behavior during training.
Thus, we can clearly see that the Markov model outranks neural network models when the LDD depth does not exceed 2, and RNN models performs the best when the depth in the dependencies does not exceed 3. 
The higher accuracy of Markov models and RNN-LSTM when the dependency depth (mutual information), exceed 2 and 3 during certain trajectory lengths could be explained by fewer number of POIs. 
We argue that the dataset containing fewer POIs and longer dependencies could be modeled by Markov processes (POI <100, MI < 2) or RNN-LSTM (POI < 100, MI >2) as they still would fit in the representational capacity of the respective models.
However for the datasets exceeding a collection duration of 2 years where the number of POIs/user > 100, and the dependency depth exceeds 3, HM-RNN models are necessary to model the intricate relationships.

In order to quantify the model complexity, we first analyze the computational efficiency to reach a stable loss according to Equation~\ref{eq:acc}.
We find that Markov models require a fraction of time need by the NN-based approaches and typically lies in the range of 1-5 seconds, followed by RNN-LSTM ($\approx9.5k$ seconds), Dilated-RNNs ($\approx11k$ seconds) and HM-RNNS ($\approx14k$ seconds).
Furthermore, we also quantify the model's memorization and representational capacity by estimating the compression ratio provided over each of the datasets in terms of bits required to represent each POI (see Figure~\ref{fig:compression_ratio}).
We observe that for each dataset, the best compression ratio is obtained by the model suggested by the meta-attributes. 
This experiment also highlights the computational complexity in terms of number of epochs to arrive at the global minima. 
\section{Conclusion}
\label{sec:conclusion}

In this paper, we have highlighted the inconsistencies and pitfalls in human mobility modeling and prediction research through a large scale systematic review. 
Through this review, we have attempted to systematize knowledge and provide guidelines towards performing credible mobility modeling research. 
We have exposed the consequences of relying on data-agnostic model selection and adopting inaccurate validation methodologies through experiments on three real-world mobility datasets. 
In order to address these problems, we have proposed four meta-attributes, that can accurately characterize a mobility dataset for selecting an appropriate modeling technique. 
Through a range of experiments, we have shown the applicability of our data-driven approach of model selection and analyzed the accuracy vs. complexity trade-offs associated with each. 
We offer the literature meta-data and the tools to the community with the hope to improve and advance the reliability of human-mobility modeling research.
In the future work, we will analyze the tradeoff between model size (depth, height, \#neurons) and memorization capacity specific to human-mobility behaviors. 
Furthermore, we also propose adoption of early-stopping in the future work to halts the model training at the {\it{best performance}} epoch, where the training loss is decreasing but the validation loss starts increasing.

{\scriptsize{

\bibliographystyle{abbrv}
\bibliography{sig-alternate-sample}  % sigproc.bib is the name of the Bibliography in this case

}}

\end{document}